\documentclass[12pt]{article}
\renewcommand{\thefootnote}{\fnsymbol{footnote}}
\newlength{\pubnumber} 

\catcode`\@=11
\@addtoreset{equation}{section}

\def\section{\@startsection{section}{1}{\z@}{3.5ex plus 1ex minus .2ex}
 {2.3ex plus .2ex}{\large\bf}}
\def\subsection{\@startsection{subsection}{2}{\z@}{2.3ex plus .2ex}
 {2.3ex plus .2ex}{\bf}}

\newcommand{\cc}[2]{c{#1\atopwithdelims[]#2}} 
\usepackage[centertags]{amsmath}
\usepackage{amssymb}
\usepackage{epsfig}
\usepackage{psfig}
\usepackage{cite}
\begin{document}
\begin{titlepage}
\samepage{
\setcounter{page}{1}
\rightline{LTH--699}
\rightline{IPPP/06/26}
\rightline{DCPT/06/52}
\rightline{\tt hep-th/0605117}
\vfill
\begin{center}
 {\Large \bf  \boldmath{${\mathbb Z}_2\times {\mathbb Z}_2$} Heterotic
 Orbifold 
 Models of  
\\
		Non Factorisable Six Dimensional Toroidal Manifolds\\}
\vfill
\vfill
 {\large Alon E. Faraggi$^{1}$\footnote{
        E-mail address: faraggi@amtp.liv.ac.uk},
         Stefan F\"orste$^{2}$\setcounter{footnote}{6}\footnote{
        E-mail address: stefan.forste@durham.ac.uk} and 
	  Cristina Timirgaziu$^{1}\setcounter{footnote}{3}$\footnote{
        E-mail address: timirgaz@amtp.liv.ac.uk}\\}	
\vspace{.12in}
 {\it $^{1}$ Department of Mathematical Sciences,
		University of Liverpool,     
                Liverpool L69 7ZL\\
and\\
	$^{2}$ Institute for Particle Physics Phenomenology,
		University of Durham,
		South Road, Durham DH1 3LE}
\end{center}
\vfill
\begin{abstract}
  {\rm
We discuss heterotic strings on ${\mathbb Z}_2 \times {\mathbb Z}_2$
orbifolds of non factorisable six-tori. Although the number of fixed
tori is 
reduced as compared to the factorisable case, Wilson lines are still
needed for the construction of three generation models. An essential
new feature is the straightforward appearance of three generation
models with one 
generation per twisted sector. We illustrate our general arguments for
the occurrence of that property by an explicit example. Our findings
give further support for the conjecture that four dimensional
heterotic strings formulated at the free fermionic point are related to
 ${\mathbb Z}_2 \times {\mathbb Z}_2$ orbifolds.}
\end{abstract}
\smallskip}
\end{titlepage}

\renewcommand{\thefootnote}{\arabic{footnote}}
\setcounter{footnote}{0}

\def\l{\label}
\def\beq{\begin{equation}}
\def\eeq{\end{equation}}
\def\beqn{\begin{eqnarray}}
\def\eeqn{\end{eqnarray}}

\def\ie{{\it i.e.}}
\def\eg{{\it e.g.}}
\def\half{{\textstyle{1\over 2}}}
\def\third{{\textstyle {1\over3}}}
\def\quarter{{\textstyle {1\over4}}}
\def\m{{\tt -}}
\def\p{{\tt +}}

\def\slash#1{#1\hskip-6pt/\hskip6pt}
\def\slk{\slash{k}}
\def\GeV{\,{\rm GeV}}
\def\TeV{\,{\rm TeV}}
\def\y{\,{\rm y}}
\def\SM{Standard-Model }
\def\SUSY{supersymmetry }
\def\SSSM{supersymmetric standard model}
\def\vev#1{\left\langle #1\right\rangle}
\def\l{\langle}
\def\r{\rangle}

\def\Htw{{\tilde H}}
\def\chibar{{\overline{\chi}}}
\def\qbar{{\overline{q}}}
\def\ibar{{\overline{\imath}}}
\def\jbar{{\overline{\jmath}}}
\def\Hbar{{\overline{H}}}
\def\Qbar{{\overline{Q}}}
\def\abar{{\overline{a}}}
\def\alphabar{{\overline{\alpha}}}
\def\betabar{{\overline{\beta}}}
\def\tautwo{{ \tau_2 }}
\def\thetatwo{{ \vartheta_2 }}
\def\thetathree{{ \vartheta_3 }}
\def\thetafour{{ \vartheta_4 }}
\def\ttwo{{\vartheta_2}}
\def\tthree{{\vartheta_3}}
\def\tfour{{\vartheta_4}}
\def\ti{{\vartheta_i}}
\def\tj{{\vartheta_j}}
\def\tk{{\vartheta_k}}
\def\calF{{\cal F}}
\def\smallmatrix#1#2#3#4{{ {{#1}~{#2}\choose{#3}~{#4}} }}
\def\ab{{\alpha\beta}}
\def\Minv{{ (M^{-1}_\ab)_{ij} }}
\def\bone{{\bf 1}}
\def\ii{{(i)}}
\def\V{{\bf V}}
\def\b{{\bf b}}
\def\N{{\bf N}}
\def\t#1#2{{ \Theta\left\lbrack \matrix{ {#1}\cr {#2}\cr }\right\rbrack }}
\def\C#1#2{{ C\left\lbrack \matrix{ {#1}\cr {#2}\cr }\right\rbrack }}
\def\tp#1#2{{ \Theta'\left\lbrack \matrix{ {#1}\cr {#2}\cr }\right\rbrack }}
\def\tpp#1#2{{ \Theta''\left\lbrack \matrix{ {#1}\cr {#2}\cr }\right\rbrack }}
\def\l{\langle}
\def\r{\rangle}


\def\inbar{\,\vrule height1.5ex width.4pt depth0pt}

\def\IC{\relax\hbox{$\inbar\kern-.3em{\rm C}$}}
\def\IQ{\relax\hbox{$\inbar\kern-.3em{\rm Q}$}}
\def\IR{\relax{\rm I\kern-.18em R}}
 \font\cmss=cmss10 \font\cmsss=cmss10 at 7pt
\def\IZ{\relax\ifmmode\mathchoice
 {\hbox{\cmss Z\kern-.4em Z}}{\hbox{\cmss Z\kern-.4em Z}}
 {\lower.9pt\hbox{\cmsss Z\kern-.4em Z}}
 {\lower1.2pt\hbox{\cmsss Z\kern-.4em Z}}\else{\cmss Z\kern-.4em Z}\fi}

\def\AEF{A.E. Faraggi}
\def\NPB#1#2#3{{Nucl.\ Phys.}\/ {B \bf #1} (#2) #3}
\def\PLB#1#2#3{{Phys.\ Lett.}\/ {B \bf #1} (#2) #3}
\def\PRD#1#2#3{{Phys.\ Rev.}\/ {D \bf #1} (#2) #3}
\def\PRL#1#2#3{{Phys.\ Rev.\ Lett.}\/ {\bf #1} (#2) #3}
\def\PRP#1#2#3{{Phys.\ Rep.}\/ {\bf#1} (#2) #3}
\def\MODA#1#2#3{{Mod.\ Phys.\ Lett.}\/ {\bf A#1} (#2) #3}
\def\IJMP#1#2#3{{Int.\ J.\ Mod.\ Phys.}\/ {A \bf #1} (#2) #3}
\def\nuvc#1#2#3{{Nuovo Cimento}\/ {\bf #1A} (#2) #3}
\def\JHEP#1#2#3{{JHEP} {\textbf #1}, (#2) #3}
\def\etal{{\it et al\/}}

\hyphenation{su-per-sym-met-ric non-su-per-sym-met-ric}
\hyphenation{space-time-super-sym-met-ric}
\hyphenation{mod-u-lar mod-u-lar--in-var-i-ant}


\setcounter{footnote}{0}
\section{Introduction}
\bigskip

String theory continues to provide the only viable framework for
the unification of all the known fundamental matter and interactions. 
Pivotal steps in this development were the discoveries of anomaly cancellation
in ten dimensions \cite{gs} and the subsequent realisation that 
compactification
to four dimensions yields the structures anticipated in
Grand Unified Theories \cite{chsw}. Further studies revealed that the 
different string theories in ten
dimensions, together with eleven dimensional supergravity, probe, in fact,
a single underlying theory \cite{dhishtw}.
The downside to these promising developments 
is that the compactification of string theory to four dimensions yields a
large number of possible vacua and, a priori, there does not exist
an apparent mechanism that selects among them. 

On the other hand, the data extracted from collider, and other, experiments
yielded the Standard Particle Model as the correct accounting of all the
observed data in the accessible energy range. Furthermore, the particle physics
data is compatible with the hypothesis that the renormalisable Standard Model
remains unaltered up to a large energy scale and that the particle spectrum
is embedded in a Grand Unified Theory \cite{gutsreviews}. 
Most appealing in this context is SO(10)
unification, in which each Standard Model generation is embedded in a 
single 16
spinorial representation of SO(10). 
However, elucidating further the properties
of the Standard Model spectrum, such as the existence of flavor, necessitates
the unification of the Standard Particle Model with gravity. 

Spinorial representations of SO(10) are obtained in the 
heterotic string \cite{hete},
but not in the other perturbative limits of the underlying 
non perturbative theory.
Thus, maintaining the SO(10) embedding of the Standard Model spectrum 
necessitates that we study compactifications of the heterotic string
on internal six dimensional manifolds. However, preserving the SO(10) 
embedding of the Standard Model spectrum in three generation string
vacua has proven to be an intricate challenge. 

A particular class of four dimensional string models, that do
yield three generations with the canonical SO(10) embedding, 
are the heterotic string models in the so--called free fermionic 
formulation \cite{fff}. The free fermionic formalism, however, is constructed
directly in four dimensions and the notion of a compactified 
manifold is a priori lost. The free fermionic models, in general, 
correspond to ${\mathbb Z}_2\times {\mathbb Z}_2$ orbifolds at special
points in the  
moduli space.
In specific cases one can make the correspondence precise \cite{foc},
while in the general case it is anticipated. 

The correspondence of the free fermionic models with ${\mathbb
Z}_2\times {\mathbb Z}_2$ 
orbifolds therefore hinges on these compactifications as the interesting
ones to explore for phenomenological purposes. Such studies were indeed 
pursued over the last few years \cite{nfw}.
However, only with partial success, in
the sense that, while three generation models were found they, 
did not yield the standard SO(10) embedding of the Standard Model
spectrum. Specifically, they did not produce models in which the
weak hypercharge has the canonical SO(10) normalisation.
The results of \cite{nfw} suggest, however, that there is a vast
number of three generation models if discrete Wilson lines
\cite{Ibanez:1986tp} are 
included into the construction. Indeed, on factorisable lattices, the
number of possible independent Wilson lines is six. The number of
models is large such that even a computer
aided classification would be a quite formidable task. More intuition
from elsewhere is needed.  

It is therefore drawn upon us to seek further insight from the
free fermionic models. The primary property of the free fermionic
formulation is of being formulated a priori
at a maximally symmetric point in the toroidal lattice
moduli space. The distinct feature is that the enhanced lattice
at the free fermionic point in the moduli space is a genuine 
$T^6$ lattice and is not factorisable to a product of three $T^2$ 
tori. The ${\mathbb Z}_2\times {\mathbb Z}_2$ heterotic orbifolds that
have been 
constructed to date are all on factorisable lattices. 

In this paper, we therefore extend these studies to ${\mathbb
Z}_2\times {\mathbb Z}_2$
orbifolds on non factorisable lattices. We elucidate further the connection
between the free fermionic models and ${\mathbb Z}_2\times {\mathbb
Z}_2$ orbifolds  
by demonstrating how the ${\mathbb Z}_2\times {\mathbb Z}_2$ fixed
point structure is modified  
on non factorisable orbifolds and, in the case of SO(12) lattice,
matches that of the free fermionic model. While this result 
has been known for some time, our discussion here is novel in the 
sense that we demonstrate how the total number of fixed points
on the SO(12) lattice is reduced due to the SO(12) root lattice
identifications. We then analyse ${\mathbb Z}_2\times {\mathbb Z}_2$
heterotic orbifolds 
on other non factorisable lattices. We include Wilson lines
in the analysis and present one three generations model in this 
class. Our paper therefore represents important advance in elucidating the
geometrical correspondence of the free fermionic models, as well
as opening the investigation of new classes of ${\mathbb Z}_2\times
{\mathbb Z}_2$ orbifolds.

The paper is organised as follows. In section two we review the free
fermionic construction, focusing especially on the features enabling a
systematic approach to the search for realistic models. The rest of
the paper is then devoted to orbifold compactifications of the
heterotic string. These construction have been developed since the 1980s,
e.g.\ in    
\cite{Dixon:1985jw,Dixon:1986jc,Ibanez:1986tp,Ibanez:1987pj,Kobayashi:1991yg,
Casas:1991ac,Giedt:2001zw}. More recent studies focus on the geometric
explanation 
of phenomenologically relevant properties
\cite{nfw,Nilles:2004ej,Forste:2005rs,Forste:2005gc,Kobayashi:2004ya,
Kobayashi:2004ud,Buchmuller:2004hv, Buchmuller:2005jr,Nilles:2006np}.

Section three addresses the geometry of ${\mathbb Z}_2 \times {\mathbb
Z}_2$ orbifolds of non factorisable tori. In particular the number of
fixed tori is explicitly counted and computed via a Lefschetz fixed
point theorem for various examples. Further, the Euler number for all
examples is determined. In section four the geometric data
are related to the number of generations for standard embeddings. As
predicted by index theorems, the net number of generations always
equals half the Euler number, whereas the number of fixed tori provides
additional information about the number of generations and
anti-generations. In section five, we use Wilson lines in order to
construct an explicit three generation model with gauge group
SO(10). In contrast to the case of a $T^6$ factorising into three
$T^2$ factors, one can easily find models with one generation per
twisted sector for non factorisable $T^6$. The reason is, that for the
latter, cycles exist which are invariant under none of the non trivial
${\mathbb Z}_2 \times {\mathbb Z}_2$ elements. More details are given
in section five. Finally, we finish with some concluding remarks in
section six.

\setcounter{footnote}{0}
\section{Realistic Free Fermionic Models}

In this section we briefly discuss the general structure of the free
fermionic models and their correspondence with ${\mathbb Z}_2\times
{\mathbb Z}_2$ heterotic 
orbifolds. The notation
and details of the construction of these
models are given elsewhere \cite{fsu5,fny,alr,euslm,nahe,cfs}.

In the free fermionic formulation, the 4-dimensional heterotic string,  in 
the
light-cone gauge, is described
by $20$ left moving  and $44$ right moving real fermions.
A large number of models can be constructed by choosing
different phases picked up by   fermions ($f_A, A=1,\dots,64$) when 
transported
along
the torus non-contractible loops.
Each model corresponds to a particular choice of fermion phases consistent 
with
modular invariance
that can be generated by a set of basis vectors $b_i,i=1,\dots,n$
$$b_i=\left\{b_i(f_1),b_i(f_{2}),b_i(f_{3})\dots\right\}$$
describing the transformation  properties of each fermion   
\begin{equation}
f_A\to -e^{i\pi b_i(f_A)}\ f_A, \ , A=1,\dots,64
\end{equation}
The basis vectors span a space $\Xi$ which consists of $2^n$ sectors that 
give
rise to the string spectrum. Each sector is given by
\begin{equation}
\alpha = \sum m_i b_i,\ \  m_i =0,\cdots,N_i
\end{equation}
The spectrum is truncated by a generalised GSO projection whose action on 
a
string state  $|S>$ is
\begin{equation}\label{eq:gso}   
e^{i\pi b_i\cdot F_S} |S> = \delta_{S}\ \cc{S}{b_i}^* |S>,
\end{equation}
where $F_S$ is the fermion number operator and $\delta_{S}=\pm1$ is the
spacetime spin statistics index.
Different sets of projection coefficients $\cc{S}{b_i}$ consistent 
with
modular invariance give
rise to different models. A model is defined uniquely by 
a set
of basis vectors $b_i,i=1,\dots,n$
and a set of $2^{n(n-1)/2}$ independent projection coefficients
$\cc{b_i}{b_j}, i>j$.

The boundary condition basis set defining a typical realistic free 
fermionic heterotic string model is constructed in two stages. The
first stage consists of the NAHE set, which is composed of five  
boundary condition basis vectors, $\{ 1 ,S,b_1,b_2,b_3\}$
\cite{costas,nahe}. The gauge group induced by the NAHE set is
${\rm SO} (10)\times {\rm SO}(6)^3\times {\rm E}_8$ with ${ N}=1$
supersymmetry. The space-time vector bosons that generate the 
gauge group arise from the Neveu--Schwarz sector and from the   
sector $\xi_2\equiv 1+b_1+b_2+b_3$. The Neveu-Schwarz sector   
produces the generators of ${\rm SO}(10)\times {\rm SO}(6)^3\times
{\rm SO}(16)$. The $\xi_2$-sector produces the spinorial 128 of
SO(16) and completes the hidden gauge group to ${\rm E}_8$. The  
NAHE set divides the internal world-sheet fermions in the
following way: ${\bar\phi}^{1,\cdots,8}$ generate the hidden ${\rm
E}_8$ gauge group, ${\bar\psi}^{1,\cdots,5}$ generate the SO(10)
gauge group, and $\{{\bar y}^{3,\cdots,6},{\bar\eta}^1\}$,
$\{{\bar y}^1,{\bar
y}^2,{\bar\omega}^5,{\bar\omega}^6,{\bar\eta}^2\}$,
$\{{\bar\omega}^{1,\cdots,4},{\bar\eta}^3\}$ generate the three
horizontal ${\rm SO}(6)$ symmetries. The left-moving 
$\{y,\omega\}$ states are divided into $\{{y}^{3,\cdots,6}\}$,
$\{{y}^1,{y}^2,{\omega}^5,{\omega}^6\}$,
$\{{\omega}^{1,\cdots,4}\}$ and $\chi^{12}$, $\chi^{34}$,
$\chi^{56}$ generate the left-moving ${ N}=2$ world-sheet
supersymmetry. At the level of the NAHE set, the sectors $b_1$,
$b_2$ and $b_3$ produce 48 multiplets, 16 from each, in the $16$
representation of SO(10). The states from the sectors $b_j$ are
singlets of the hidden ${\rm E}_8$ gauge group and transform under
the horizontal ${\rm SO}(6)_j$ $(j=1,2,3)$ symmetries. 
This
structure is common to a large class of quasi--realistic free fermionic 
models.

The second stage of the
construction consists of adding to the
NAHE set three (or four) additional basis vectors.
These additional vectors reduce the number of generations
to three, one from each of the sectors $b_1$,
$b_2$ and $b_3$, and simultaneously break the four dimensional 
gauge group. The assignment of boundary conditions to
$\{{\bar\psi}^{1,\cdots,5}\}$ breaks SO(10) to one of its subgroups
${\rm SU}(5)\times {\rm U}(1)$ \cite{fsu5}, ${\rm SO}(6)\times {\rm 
SO}(4)$
\cite{alr},
${\rm SU}(3)\times {\rm SU}(2)\times {\rm U}(1)^2$ \cite{fny,euslm},
${\rm SU}(3)\times {\rm SU}(2)^2\times {\rm U}(1)$ \cite{cfs} or
${\rm SU}(4)\times {\rm SU}(2)\times {\rm U}(1)$ \cite{cfnooij}.
Similarly, the hidden ${\rm E}_8$ symmetry is broken to one of its
subgroups, and the flavor ${\rm SO}(6)^3$ symmetries are broken   
to $U(1)^n$, with $3\le n\le9$.
For details and phenomenological studies of
these three generation string models we refer interested
readers to the original literature and review articles
\cite{reviewsp}.

The correspondence of the free fermionic models
with the orbifold construction is illustrated
by extending the NAHE set, $\{ 1,S,b_1,b_2,b_3\}$, by at least 
one additional boundary condition basis vector \cite{foc}
\beq
\xi_1=(0,\cdots,0\vert{\underbrace{1,\cdots,1}_{{\bar\psi^{1,\cdots,5}},   
{\bar\eta^{1,2,3}}}},0,\cdots,0)~.
\label{vectorx}
\eeq
With a suitable choice of the GSO projection coefficients the   
model possesses an ${\rm SO}(4)^3\times {\rm E}_6\times {\rm U}(1)^2
\times {\rm E}_8$ gauge group
and ${ N}=1$ space-time supersymmetry. The matter fields
include 24 generations in the 27 representation of
${\rm E}_6$, eight from each of the sectors $b_1\oplus b_1+\xi_1$,
$b_2\oplus b_2+\xi_1$ and $b_3\oplus b_3+\xi_1$.
Three additional 27 and $\overline{27}$ pairs are obtained
from the Neveu-Schwarz $\oplus~\xi_1$ sector.

To construct the model in the orbifold formulation one starts
with the compactification on a torus with nontrivial background
fields \cite{Narain}.
The subset of basis vectors
\beq
\{ 1,S,\xi_1,\xi_2\}
\label{neq4set}
\eeq
generates a toroidally-compactified model with ${ N}=4$ space-time
supersymmetry and ${\rm SO}(12)\times {\rm E}_8\times {\rm E}_8$ gauge 
group.
The same model is obtained in the geometric (bosonic) language
by tuning the background fields to the values corresponding to
the SO(12) lattice. The
metric of the six-dimensional compactified
manifold is then the Cartan matrix of SO(12),
while the antisymmetric tensor is given by
\begin{equation}
b_{ij}=\begin{cases}
g_{ij}&;\ i>j,\cr
0&;\ i=j,\cr
-g_{ij}&;\ i<j.\cr\end{cases}
\label{bso12}
\end{equation}
When all the radii of the six-dimensional compactified
manifold are fixed at $R_I=\sqrt2$, it is seen that the
left- and right-moving momenta  
\beq
P^I_{R,L}=[m_i-{\frac{1}{2}}(B_{ij}{\pm}G_{ij})n_j]{e_i^I}^*
\label{lrmomenta}
\eeq
reproduce the massless root vectors in the lattice of
SO(12). Here $e^i=\{e_i^I\}$ are six linearly-independent
vielbeins normalised so that $(e_i)^2=2$.
The ${e_i^I}^*$ are dual to the $e_i$, with
$e_i^*\cdot e_j=\delta_{ij}$.

Adding the two basis vectors $b_1$ and $b_2$ to the set
(\ref{neq4set}) corresponds to the ${\mathbb Z}_2\times {\mathbb Z}_2$
orbifold model with standard embedding.
Starting from the $N=4$ model with ${\rm SO}(12)\times
{\rm E}_8\times {\rm E}_8$
symmetry, and applying the ${\mathbb Z}_2\times {\mathbb Z}_2$
twist on the
internal coordinates, reproduces
the spectrum of the free-fermion model
with the six-dimensional basis set
$\{ 1,S,\xi_1,\xi_2,b_1,b_2\}$ \cite{foc}.
The Euler characteristic of this model is 48 with $h_{11}=27$ and
$h_{21}=3$.

It is noted that the effect of the additional basis vector $\xi_1$ of eq.\
(\ref{vectorx}), is to separate the gauge degrees of freedom, spanned by
the world-sheet fermions $\{{\bar\psi}^{1,\cdots,5},
{\bar\eta}^{1},{\bar\eta}^{2},{\bar\eta}^{3},{\bar\phi}^{1,\cdots,8}\}$,
from the internal compactified degrees of freedom $\{y,\omega\vert
{\bar y},{\bar\omega}\}^{1,\cdots,6}$.
In the realistic free fermionic
models this is achieved by the vector $2\gamma$ \cite{foc}, with
\beq
2\gamma=(0,\cdots,0\vert{\underbrace{1,\cdots,1}_{{\bar\psi^{1,\cdots,5}},
{\bar\eta^{1,2,3}} {\bar\phi}^{1,\cdots,4}} },0,\cdots,0)~,
\label{vector2gamma}
\eeq
which breaks the ${\rm E}_8\times {\rm E}_8$ symmetry to ${\rm 
SO}(16)\times
{\rm SO}(16)$.
The ${Z}_2\times {Z}_2$ twist induced by $b_1$ and $b_2$
breaks the gauge symmetry to
${\rm SO}(4)^3\times {\rm SO}(10)\times {\rm U}(1)^3\times {\rm SO}(16)$.
The orbifold still yields a model with 24 generations,
eight from each twisted sector,
but now the generations are in the chiral 16 representation
of SO(10), rather than in the 27 of ${\rm E}_6$. The same model can
be realised \cite{phaseq} with the set
$\{ 1,S,\xi_1,\xi_2,b_1,b_2\}$,
by projecting out the $16\oplus{\overline{16}}$
from the $\xi_1$-sector taking
\beq\label{changec}
\cc{\xi_1}{\xi_2} \rightarrow -\cc{\xi_1}{\xi_2}.
\eeq
This choice also projects out the massless vector bosons in the
128 of SO(16) in the hidden-sector ${\rm E}_8$ gauge group, thereby
breaking the ${\rm E}_6\times {\rm E}_8$ symmetry to
${\rm SO}(10)\times {\rm U}(1)\times {\rm SO}(16)$.
We can define two ${ N}=4$ models generated by the set
(\ref{neq4set}), ${ Z}_+$ and ${ Z}_-$, depending on the sign
in eq.\ (\ref{changec}). The first, say ${ Z}_+$,
produces the ${\rm E}_8\times {\rm E}_8$ model, whereas the second, say
${ Z}_-$, produces the ${\rm SO}(16)\times {\rm SO}(16)$ model.
However, the ${\mathbb Z}_2\times
{\mathbb Z}_2$
twist acts identically in the two models, and their physical 
characteristics
differ only due to the discrete torsion eq.\ (\ref{changec}).

Several remarks are important to note at this stage. The first is
that a priori the point at which the internal dimensions are realised as
free fermions on the world--sheet is a maximally symmetric point with 
an enhanced $SO(12)$ lattice. This lattice is a priori not factorisable
to a product of three $T^2$ lattices, but is rather a non factorisable 
$T^6$ lattice. The phenomenologically appealing properties of the
free fermionic models and their relation to ${\mathbb Z}_2\times
{\mathbb Z}_2$ orbifolds 
provide the clue that we might gain further insight into
the properties of this class of quasi--realistic string compactifications 
by constructing
${\mathbb Z}_2\times {\mathbb Z}_2$ orbifolds on enhanced
non factorisable lattices.  

It is further important to note that the free fermionic formalism enables
a classification of a wide range of ${\mathbb Z}_2\times {\mathbb
Z}_2$ heterotic orbifolds. 
The correspondence illustrated above exhibits in detail the
correspondence 
of a single free fermionic model with a ${\mathbb Z}_2\times {\mathbb
Z}_2$ orbifold in a 
specific case. The correspondence in the general case is seen
by analysing the respective partition functions and it is anticipated that for every ${\mathbb Z}_2\times {\mathbb
Z}_2$ orbifold 
one can write a corresponding free fermionic partition function
at special points in the moduli space, and hence obtain a representation
in terms of free fermion boundary condition basis vectors.
In turn, using the free fermion formalism the classification 
of ${\mathbb Z}_2\times {\mathbb Z}_2$ orbifolds is carried out by
specifying a set  
of boundary condition basis vectors and one--loop GSO coefficients. 
The details of this classification for type II strings were carried
in \cite{gkr} and for the heterotic string in \cite{fknr}.

The free fermionic formalism provides useful means to classify and
analyse ${\mathbb Z}_2\times {\mathbb Z}_2$ heterotic orbifolds at
special points in the  
moduli space. The drawbacks of this approach is that the analysis 
is carried out at special points in the moduli space and the geometric
view of the underlying compactifications is hindered. On the other 
hand, the geometric picture may be instrumental for examining other
questions of interest, such as the dynamical stabilisation of the moduli
fields and the moduli dependence of the Yukawa couplings.
In the following we turn to analyse ${\mathbb Z}_2\times {\mathbb
Z}_2$ orbifolds  
on non factorisable toroidal manifolds. 

\section{\boldmath{${\mathbb Z}_2\times {\mathbb Z}_2$} Orbifolds of
Non Factorisable Tori} 

In this section we discuss the ${\mathbb Z}_2\times {\mathbb Z}_2$
orbifold of 
non factorisable $T^6$  
tori. We use a geometrically intuitive picture that facilitates 
understanding the fixed point structure on these lattices. We pay
special attention to  
the identifications by the non factorisable root lattice vectors
and how the number of fixed tori is affected. Our discussion
illuminates a long standing puzzle in the orbifold community 
in regard to the ${\mathbb Z}_2\times {\mathbb Z}_2$ orbifold
correspondence of  
the free fermionic models. 

In the following we will exemplarily discuss a set of
compactifications on non factorisable six-tori. These will illustrate
the main features of such compactifications. For more systematic
studies (from slightly different perspectives) see
\cite{foc,Dixon:1986jc,Dixon:1986yc}. We will determine the number of
fixed tori once by explicit counting and once via a Lefschetz fixed
point theorem. In addition, we give the Euler number of the
orbifold. For the standard embedding of the orbifold action into the
gauge group the net number of generations will be fixed by the Euler
number, whereas the number of generations and anti-generations depends
also on the number of fixed tori. (For different orbifolds this
is discussed in \cite{Erler:1992ki}.) 

We specify the ${\mathbb Z}_2 \times {\mathbb Z}_2$ orbifold action on
a set of six Cartesian coordinates  $x^1, \ldots , x^6$ of the compact
space as follows: 
\begin{equation}\label{eq:orbiact}
\left( \begin{array}{c}  x^1 \\ \vdots \\ x^6 \end{array}\right)
\rightarrow \theta_1 \left( \begin{array}{c}  x^1 \\ \vdots \\ x^6
\end{array}\right) , \,\,\, \mbox{with} \,\,\, \theta_1 = \left(
\begin{array}{ c c c c c c}
-1     &  0   &   0  &  0  & 0 &  0 \\
0     &  -1   &   0  &  0 &  0 & 0  \\
0     &  0   &   -1 &  0 &  0 & 0 \\
0    &   0   &   0  &  -1 & 0 & 0 \\
0   &    0   &   0  &   0 & 1& 0 \\
0   &   0 &  0 &  0  & 0 & 1  \end{array} \right) 
\end{equation}
and
\begin{equation}\label{eq:orbiact2}
\left( \begin{array}{c}  x^1 \\ \vdots \\ x^6 \end{array}\right)
\rightarrow \theta_2 \left( \begin{array}{c}  x^1 \\ \vdots \\ x^6
\end{array}\right) , \,\,\, \mbox{with} \,\,\, \theta_2 = \left(
\begin{array}{ c c c c c c}
1     &  0   &   0  &  0  & 0 &  0 \\
0     &  1   &   0  &  0 &  0 & 0  \\
0     &  0   &   -1 &  0 &  0 & 0 \\
0    &   0   &   0  &  -1 & 0 & 0 \\
0   &    0   &   0  &   0 & -1& 0 \\
0   &   0 &  0 &  0  & 0 & -1  \end{array} \right)  ,
\end{equation}
where $\theta_1$ and $\theta_2$ are the generators of ${\mathbb Z}_2
\times {\mathbb Z}_2$.

\subsection{SO(12) Lattice}\label{sec:so12}

As a first example, we consider the case that $T^6$ is obtained by
compactifying ${\mathbb R}^6$ on an SO(12) root lattice whose basis
vectors are given by the 
simple roots 
\begin{eqnarray}
e_1 & = & \left( 1 , -1 , 0,0,0,0\right) , \nonumber \\
e_2 & = & \left( 0, 1, -1, 0,0,0\right) , \nonumber \\
e_3 & = & \left( 0 ,0 , 1, -1,0,0\right) , \nonumber \\
e_4 & = & \left( 0, 0, 0, 1,-1,0\right) , \nonumber \\
e_5 & = & \left( 0,0,0,0,1,-1\right) , \nonumber \\
e_6 & = & \left( 0,0,0,0,1,1\right) .\label{eq:so12roots}
\end{eqnarray}

The orbifold action, by e.g.\ (\ref{eq:orbiact2}), leaves sets of points
invariant, i.e.\ these points differ from their orbifold image by an
SO(12) root lattice shift. For our particular choice of the orbifold
action these sets appear as two dimensional fixed tori. In the
following we will list 16 such two-tori and afterwards argue that some
of these 16 tori are identical. That will leave us in the end with
eight different two-tori. 

The trivial fixed torus is given as the set
\begin{equation}\label{eq:fix1}
\left\{ \left( x , y , 0,0,0,0\right) \left|\, x,y \in {\mathbb R}^2/
\Lambda^2 \right. \right\} .
\end{equation}
The compactification lattice $\Lambda^2$ is generated by the vectors
$\left( 1 ,1 \right)$ and $\left( 1 , -1 \right)$. This can be
verified by writing
$$ \left( x,y,0,0,0,0\right) = x e_1 + \left( x+y\right) \left(e_2 + e_3 +
e_4 + \frac{1}{2} e_5 + \frac{1}{2} e_6 \right) $$
and identifying minimal shifts in $(x,y)$ shifting the coefficients in
front of lattice vectors by integers.

Now, consider the fixed torus 
\begin{equation}\label{eq:fix2}
\left\{ \left( x , y , 1,0,0,0\right) \left|\,  x,y \in {\mathbb R}^2/
\Lambda^2 \right. \right\} .
\end{equation}
Points on that torus differ from their image point by the lattice
vector $\left( 0,0,2,0,0,0\right)$. The position of the 1
entry can be altered within the last four components by adding SO(12)
root vectors, e.g.\ $\left( 0,0,-1,1,0,0\right)$.  

Next there are fixed tori of the form
\begin{equation}\label{eq:fix3}
\left\{ \left( x , y , \underline{\frac{1}{2},\frac{1}{2},0,0}\right)
  \left|\, x,y \in {\mathbb R}^2/ 
\Lambda^2 \right. \right\} ,
\end{equation}
where the underlined entries can be permuted. Points on these fixed
tori differ from their orbifold image by an SO(12) root, e.g.\ 
$\left( 0 ,0 , 1, 1, 0,0\right)$. There are 
$$ \left( \begin{array}{c} 4 \\  2 \end{array} \right) = 6 $$
such fixed two-tori. 

Very similar fixed tori are

\begin{equation}\label{eq:fix4}
\left\{ \left( x , y , \underline{\frac{1}{2},-\frac{1}{2},0,0}\right)
  \left|\, x,y \in {\mathbb R}^2/ 
\Lambda^2 \right. \right\} ,
\end{equation}
where the position of the minus sign can be changed by lattice shifts
( $\left(1/2, -1/2\right) + \left( -1, 1\right) = \left( -1/2,
1/2\right)$).
This yields another set of six fixed tori.

Finally, there are the fixed tori 
\begin{equation}\label{eq:fix5}
\left\{ \left( x , y ,
  \frac{1}{2},\frac{1}{2},\frac{1}{2},\frac{1}{2}\right) 
  \left|\, x,y \in {\mathbb R}^2/ 
\Lambda^2 \right. \right\} 
\end{equation}
and
\begin{equation}\label{eq:fix6}
\left\{ \left( x , y ,
  \frac{1}{2},\frac{1}{2},\frac{1}{2},-\frac{1}{2}\right) 
  \left|\, x,y \in {\mathbb R}^2/ 
\Lambda^2 \right. \right\} .
\end{equation}

So, altogether we have listed 16 fixed tori. Now, we are going to
argue that some of them are equivalent. Consider the fixed torus
(\ref{eq:fix2}) and add the SO(12) root vector $\left(
1,0,-1,0,0,0\right)$. This yields an equivalent expression for
(\ref{eq:fix2}) 
\begin{equation}
\left\{ \left( x+1 , y ,
  0,0,0,0\right) 
  \left|\, x,y \in {\mathbb R}^2/ 
\Lambda^2 \right. \right\} .
\end{equation}

But this is the same fixed torus as (\ref{eq:fix1}), merely the origin
for the $x$ coordinate has been shifted by one. Similar arguments show
that the tori in (\ref{eq:fix3}) and (\ref{eq:fix4}) as well as the
tori (\ref{eq:fix5}) and (\ref{eq:fix6}) are mutually equivalent. So,
finally we are left with eight inequivalent fixed tori.

For the ${\mathbb Z}_2 \times {\mathbb Z}_2$ orbifold we add another
${\mathbb Z}_2$ action $\theta_1$ (\ref{eq:orbiact}).
For this ${\mathbb Z}_2 \times {\mathbb Z}_2$ action we obtain eight fixed
tori under the action of $\theta_1$, eight fixed tori under the action
of $\theta_2$ 
and eight fixed tori under the action of $\theta_1\theta_2$. Hence, the
total number 
of fixed tori is 24.

After the somewhat tricky explicit counting we would like to confirm
our result by using mathematical fixed point theorems. For the case of
fixed tori the adequate Lefschetz fixed point theorem for the number
of fixed tori (\# FT) reads \cite{Narain:1986qm}\footnote{In our
coordinate system the orbifold action is represented by symmetric
matrices and in that case (\ref{eq:lefschetz}) is equivalent to the
expression given in \cite{akin} where a different calculation is described.}
\begin{equation}\label{eq:lefschetz}
\# FT = \left| \frac{N}{\left( 1 - \theta\right) \Lambda}\right| .
\end{equation}
On the rhs the index of a lattice quotient appears. The lattice in the
denominator is a sub-lattice of the lattice in the numerator and the
index counts how often the fundamental cell of the finer
lattice fits into the one of the
coarser one.  In our case 
the coarser lattice is $\left( 1 - \theta\right) \Lambda$, which is a
sub-lattice of the compactification lattice $\Lambda$, obtained by
projecting with $( 1 -\theta )$. The finer lattice is $N$, which is
the lattice normal to the sub-lattice left invariant by the action of
$\theta$. A more handy version of (\ref{eq:lefschetz}) is
\begin{equation}\label{eq:lefschetzvol}
\# FT =  \frac{\mbox{vol}\left(\left( 1 - \theta\right)
  \Lambda\right)}{\mbox{vol}\left( N\right)} , 
\end{equation}
where the volumes of the fundamental cells appear on the rhs. These can
be easily determined by computing the induced metrics on the
corresponding lattices. 
Now let us illustrate the theorem at the case of $\theta_2$ fixed tori
in the SO(12) lattice. First, we note that
\begin{equation}
1- \theta_2 = \left(
\begin{array}{ c c c c c c}
0     &  0   &   0  &  0  & 0 &  0 \\
0     &  0   &   0  &  0 &  0 & 0  \\
0     &  0   &   2 &  0 &  0 & 0 \\
0    &   0   &   0  &  2 & 0 & 0 \\
0   &    0   &   0  &   0 & 2& 0 \\
0   &   0 &  0 &  0  & 0 & 2  \end{array} \right) .
\end{equation}
A basis in the $\left( 1 - \theta_2\right)
\Lambda$ lattice is
\begin{equation}
\left( 1 - \theta_2\right) \Lambda: \,\,\, \begin{array}{c}
  \left( 0,0,2,0,0,0\right) , \\ \left( 0,0,0,2,0,0\right) , \\
\left( 0,0,0,0,2,0\right) ,\\ \left( 0,0,0,0,0,2\right) , \end{array} 
\end{equation}
where e.g.\ the first basis vector originates from the SO(12) root
$\left( 1,0,1,0,0,0\right)$. Hence, the induced metric is four times a
four by four unit matrix and the volume of the fundamental cell can
be obtained as the square root of its determinant
\begin{equation}\label{eq:SO12L}
\mbox{vol}\left( \left( 1 - \theta_2\right) \Lambda\right) = 16 .
\end{equation}
The $\theta_2$ invariant lattice is spanned by $\left(
1,1,0,0,0,0\right)$ and $\left( 1,-1,0,0,0,0\right)$. Hence, the
normal lattice is generated by simple SO(8) roots
\begin{equation}
N:\,\,\, \begin{array}{c} \left( 0,0,1,-1,0,0\right) ,\\
\left( 0,0,0,1,-1,0\right) , \\ \left( 0,0,0,1, -1\right) , \\
\left( 0,0,0,0,1,1\right) .
\end{array}
\end{equation}
The induced metric is the SO(8) Cartan matrix
\begin{equation}
\left( \begin{array}{ rrrr }
 2 & -1 & 0 & 0 \\
-1 & 2 & -1 & -1 \\
0 & -1 & 2 & 0\\
0 & -1 & 0 & 2
\end{array} \right) .
\end{equation}
The square root of the determinant of the induced metric provides the
volume 
\begin{equation}\label{eq:SO12N}
\mbox{vol}\left( N \right) = 2 .
\end{equation}
Plugging (\ref{eq:SO12L}) and (\ref{eq:SO12N}) into
(\ref{eq:lefschetzvol}), we find that the number of $\theta_2$ fixed
tori is eight, confirming our earlier result, obtained by explicit
counting. The calculation for the other ${\mathbb Z}_2 \times
{\mathbb Z}_2$ non trivial elements $\theta_1$ and $\theta_1 \theta_2$
is a straightforward modification of the presented calculation. Each
element leaves eight tori fixed, yielding a total of 24 fixed tori.

Finally, we would like to compute the Euler number of the given
orbifold. The general expression for the Euler number $\chi$  is
\cite{Dixon:1985jw} 
\begin{equation}
\chi = \frac{1}{\left| G \right|} \sum_{[g,h] = 0} \chi_{g,h} ,
\end{equation}
where $\left| G \right|$ is the order of the orbifold group with
elements $g$, $h$ and $\chi_{g,h}$ is the number of points which are
simultaneously fixed under the action of $g$ and $h$.
(Here, it is important that these are really points, if e.g.\ $g=1$
then $\chi_{1,h}$ provides the number of tori fixed under $h$, which do
not contribute to the Euler number.)

Now, let us identify points which are fixed under $\theta_1$ and
$\theta_2$ in the case that the compactification is on the SO(12) root
lattice. One finds 32 such points:
\begin{equation}
\begin{array} {l l l}
\left( 0,0,0,0,0,0\right) & & \mbox{1 point}, \\
\left( 1 ,0,0,0,0,0\right) & & \mbox{ 1 point}, \\
\left( \frac{1}{2}, \pm \frac{1}{2}, 0,0, 0,0\right) & + S_3 & \mbox{6
  points},\\
\left( \frac{1}{2}, \frac{1}{2}, \frac{1}{2}, \pm \frac{1}{2},
0,0\right) & + S_3 & \mbox{6 points},\\
\left( \frac{1}{2}, \frac{1}{2}, \frac{1}{2}, \frac{1}{2},\frac{1}{2},
\pm \frac{1}{2} \right) & & \mbox{2 points},\\
\left( \underline{\frac{1}{2}, 0}, \underline{\frac{1}{2}, 0},
  \underline{\pm \frac{1}{2}, 0}\right) & & \mbox{16 points},
\end{array}
\end{equation}
where ``$+S_3$'' stands for adding vectors obtained by permuting the
first, second and third pair of entries and underlined entries can
also be permuted. So, altogether, we find
\begin{equation}
\chi_{\theta_1, \theta_2} = 32.
\end{equation}
This is four times the number of tori fixed under $\theta_2$.  
The counting for the other five combinations of non-trivial ${\mathbb
  Z}_2 \times {\mathbb Z}_2$ elements is the same and we end up with
\begin{equation}
\chi = 48, 
\end{equation}
which is twice the number of fixed tori.

\subsection{SO(6)$^2$ Lattice - First Example}\label{sec:so6-1}

\vspace*{.3in}

Now, we choose for the $T^6$ compactification an SO(6)$^2$ lattice
with basis vectors 
\begin{eqnarray}
e_1 & = & \left( 1 , -1 , 0,0,0,0\right) , \nonumber \\
e_2 & = & \left( 0, 1, -1, 0,0,0\right) , \nonumber \\
e_3 & = & \left( 0 ,1, 1, 0,0,0\right) , \nonumber \\
e_4 & = & \left( 0, 0, 0, 1,-1,0\right) , \nonumber \\
e_5 & = & \left( 0,0,0,0,1,-1\right) , \nonumber \\
e_6 & = & \left( 0,0,0,0,1,1\right) .\label{eq:simpso61}
\end{eqnarray}

The determination of the fixed tori under $\theta_2$ is very similar
to the previous 
case. First, we point out the differences and then list the fixed
tori. The fixed torus
\begin{equation}\label{eq:newfixed}
\left\{ \left( x , y ,
  0,0,0,1\right) 
  \left|\, x,y \in {\mathbb R}^2/ 
\Lambda^2 \right. \right\} ,
\end{equation}
is not equivalent to
\begin{equation}\label{eq:trivialfixed}
\left\{ \left( x, y ,
  0,0,0,0\right) 
  \left|\, x,y \in {\mathbb R}^2/ 
\Lambda^2 \right. \right\} ,
\end{equation}
since there is no lattice vector like $\left( 1,0,0,0,0,-1\right)$ in
the SO(6)$^2$ lattice. On the other hand, the fixed torus
\begin{equation}
\left\{ \left( x , y ,
  1,0,0,0\right) 
  \left|\, x,y \in {\mathbb R}^2/ 
\Lambda^2 \right. \right\} 
\end{equation}
is equivalent to (\ref{eq:trivialfixed}) (via a shift with $\left(
1,0,-1,0,0,0\right)$ and a reparameterisation of the fixed torus). It
is not equivalent to (\ref{eq:newfixed}) because the SO(12) lattice
vector $\left( 0,0,-1,0,0,1\right)$ is not in the SO(6)$^2$ lattice. 
Taking this kind of arguments into account one finds the following
eight fixed tori under the action of $\theta_2$ (\ref{eq:orbiact2})
\begin{eqnarray}
\left\{ \left( x , y ,
  0,0,0,0\right) 
  \left|\, x,y \in {\mathbb R}^2/ 
\Lambda^2 \right. \right\} , & & \mbox{1 fixed torus},\label{eq:so6fixf}\\
\left\{ \left( x , y ,
  0,0,0,1\right) 
  \left|\, x,y \in {\mathbb R}^2/ 
\Lambda^2 \right. \right\} , & & \mbox{1 fixed
  torus},\label{eq:sector} \\ 
\left\{ \left( x , y ,
  0,\underline{\frac{1}{2},\frac{1}{2},0}\right)
  \left|\, x,y \in {\mathbb R}^2/ 
\Lambda^2 \right. \right\} , & & \mbox{3 fixed
  tori}, \label{eq:thirtor}\\ 
\left\{ \left( x , y ,
  0,\underline{\frac{1}{2},-\frac{1}{2},0}\right)
  \left|\, x,y \in {\mathbb R}^2/ 
\Lambda^2 \right. \right\} , & & \mbox{3 fixed
  tori},\label{eq:minusnomat}
\end{eqnarray}
where underlined entries can be again permuted and the position of the
minus sign within the last three entries in (\ref{eq:minusnomat}) can
be altered by lattice shifts.

By swapping the two SO(6) factors, we obtain the following eight fixed
tori under the action of $\theta_1$ (\ref{eq:orbiact})
\begin{eqnarray}
\left\{ \left( 0 , 0 ,
  0,0,x,y\right) 
  \left|\, x,y \in {\mathbb R}^2/ 
\Lambda^2 \right. \right\} , & & \mbox{1 fixed torus},\\
\left\{ \left( 
  1,0,0,0,x,y\right) 
  \left|\, x,y \in {\mathbb R}^2/ 
\Lambda^2 \right. \right\} , & & \mbox{1 fixed torus}, \\
\left\{ \left( \underline{\frac{1}{2},\frac{1}{2},0},0,x,y\right)
  \left|\, x,y \in {\mathbb R}^2/ 
\Lambda^2 \right. \right\} , & & \mbox{3 fixed
  tori}, \\ 
\left\{ \left(\underline{\frac{1}{2},-\frac{1}{2},0},0,x,y\right)
  \left|\, x,y \in {\mathbb R}^2/ 
\Lambda^2 \right. \right\} , & & \mbox{3 fixed
  tori}.
\end{eqnarray}
For the action of $\theta_1 \theta_2$ the situation is different. As a
first difference, we note that the trivially fixed torus is
\begin{equation}
\left\{ \left( 0 , 0 ,
  x,y,0,0\right) 
  \left|\, x,y \in {\mathbb R}^2/ 
\tilde{\Lambda}^2 \right. \right\},
\end{equation}
with a modified lattice $\tilde {\Lambda}^2$ which is generated by
$(2,0)$ and $(0,2)$. (The point $(0,0,x,y,0,0) = \frac{x}{2} \left( 
e_2 + e_3\right) +\frac{y}{2}\left( 2e_4 + e_5 + e_6\right)$ on $T^6$
is invariant if $x$ or $y$ are shifted by integer multiples of two.) This
fixed torus is twice as big as the ones which occurred previously.  
Now the fixed tori 
$$\left\{ \left( 1, 0 ,
  x,y,0,0\right) 
  \left|\, x,y \in {\mathbb R}^2/ \tilde{\Lambda}^2 \right. \right\}$$
as well as
$$\left\{ \left( 0 , 0 ,
  x,y,1,0\right) 
  \left|\, x,y \in {\mathbb R}^2/ 
\tilde{\Lambda}^2 \right. \right\}$$ are equivalent to the trivial one
(by shifts with $(1,0,-1,0,0,0)$ or $(0,0,0,1,-1,0)$ and redefinitions
of the torus coordinates $x$ or $y$). 
One finds four inequivalent fixed tori:
\begin{eqnarray}
\left\{ \left( 0 , 0 , x,y,0,0\right) 
  \left|\, x,y \in {\mathbb R}^2/ 
\tilde{\Lambda}^2 \right. \right\} , & & \mbox{1 fixed
  torus},\label{eq:onetwof}\\ 
\left\{ \left( 
  \frac{1}{2},\frac{1}{2},x,y,0,0\right) 
  \left|\, x,y \in {\mathbb R}^2/ 
\tilde{\Lambda}^2 \right. \right\} , & & \mbox{1 fixed torus}, \\
\left\{ \left( 0,0,x,y, \frac{1}{2}, \frac{1}{2}\right)
  \left|\, x,y \in {\mathbb R}^2/ 
\tilde{\Lambda}^2 \right. \right\} , & & \mbox{1 fixed
  torus}, \\ 
\left\{ \left(\frac{1}{2},\frac{1}{2},x,y,\frac{1}{2},\frac{1}{2}\right)
  \left|\, x,y \in {\mathbb R}^2/ 
\tilde{\Lambda}^2 \right. \right\} , & & \mbox{1 fixed
  torus}.\label{eq:onetwoff}
\end{eqnarray}
Hence, we obtain 20 fixed tori in total (eight under the action of
$\theta_1$ and $\theta_2$ each and another four from the action of
$\theta_1 \theta_2$). 

Before rederiving this result using the Lefschetz theorem, let us
comment on what we found so 
far. The number of fixed tori under the action of $\theta_1$ and
$\theta_2$ is the same as for the SO(12) lattice. In fact, as long as we
project only on the e.g.\ $\theta_2$ invariant states, moduli allowing
a deformation into the SO(12) lattice are still present in the
spectrum. In particular, the internal metric components, $g_{IJ}$, with
$I,J \geq 3$ are even under the action of $\theta_1$. The
compactification lattice can be deformed within the $x^3$-$x^4$-$x^5$-$x^6$
`plane'. This means that we can continuously deform $e_6$ in
(\ref{eq:simpso61} ) into $( 0,0,1,1,0,0)$ yielding an SO(12)
lattice. Likewise, under $\theta_1$, moduli allowing a deformation
within the $x^1$-$x^2$-$x^3$-$x^4$ `plane' survive the
projection. Replacing $e_2$ with $(0,0,1,-1,0,0)$ can be done in a
continuous way, yielding again an SO(12) lattice. On the other hand,
moduli which are even under 
$\theta_1 \theta_2$ allow for deformations within the 
$x^1$-$x^2$-$x^5$-$x^6$ and $x^3$-$x^4$ planes. Neither of the above
replacements 
yielding the SO(12) lattice corresponds to such a deformation. Thus
our findings for the numbers of fixed tori in the different sectors
are compatible with possible continuous deformations of the orbifold.   

Now, we confirm the result obtained by explicit counting via the
Lefschetz fixed point theorem (\ref{eq:lefschetz}). 
First, we compute the number of fixed tori under $\theta_2$. A basis
for the $\left( 1 - \theta_2\right) \Lambda $ lattice is
\begin{equation}
\left( 1 - \theta_2\right) \Lambda: \,\,\, 
\begin{array}{c} \left(0,0,2,0,0,0\right) ,\\ 
\left( 0,0,0,2,-2,0\right), \\
\left( 0,0,0,0,2,-2\right) , \\
\left( 0,0,0,0,2,2\right) .\end{array}
\end{equation}
The induced metric reads
\begin{equation}
4\left( \begin{array}{cccc}
1  & 0 & 0 & 0\\
0  & 2 & -1 & -1 \\
0 & -1 & 2 & 0 \\
0 & -1 & 0 & 2 
\end{array} \right) .
\end{equation}
The square root of its determinant is 32.
On the other hand the lattice normal to the invariant lattice is
\begin{equation}
N : \,\,\, 
\begin{array}{c} \left(0,0,2,0,0,0\right) ,\\ 
\left( 0,0,0,1,-1,0\right), \\
\left( 0,0,0,0,1,-1\right) , \\
\left( 0,0,0,0,1,1\right) ,\end{array}
\end{equation}
resulting in the induced metric
\begin{equation}
\left( \begin{array}{cccc}
4  & 0 & 0 & 0\\
0  & 2 & -1 & -1 \\
0 & -1 & 2 & 0 \\
0 & -1 & 0 & 2 
\end{array} \right) ,
\end{equation}
whose determinant is $4^2$. Hence, the number of tori fixed
under $\theta_2$ is 32/4 =8.

For the number of fixed tori under $\theta_1$ a completely analogous
computation yields again eight.

Now, consider the ${\mathbb Z}_2 \times {\mathbb Z}_2$ generator
$\theta_1 \theta_2$ for which
\begin{equation}
\left( 1 - \theta_1 \theta_2\right) = \left(
\begin{array}{cccccc}
2 & 0 & 0 & 0 & 0 & 0 \\
0 & 2 & 0 & 0 & 0 & 0 \\
0 & 0 & 0 & 0 & 0 & 0 \\
0 & 0 & 0 & 0 & 0 & 0 \\
0 & 0 & 0 & 0 & 2 & 0 \\
0 & 0 & 0 & 0 & 0 & 2 
\end{array}\right) .
\end{equation}
The lattice $\left( 1 - \theta_1 \theta_2\right)\Lambda$  is spanned
by the following basis
\begin{equation}
\left( 1 - \theta_1\theta_2\right) \Lambda: \,\,\, 
\begin{array}{c} \left(2,0,0,0,0,0\right) ,\\ 
\left( 0,2,0,0,0,0\right) , \\
\left( 0,0,0,0,2,0\right) , \\
\left( 0,0,0,0,0,2\right) .
\end{array}
\end{equation}
Hence, the induced metric is just four times a four by four identity
matrix. The square root of its determinant being 16.

For the normal lattice we find
\begin{equation}
N: \,\,\, 
\begin{array}{c} \left(1,-1,0,0,0,0\right) ,\\ 
\left( 1,1,0,0,0,0\right), \\
\left( 0,0,0,0,1,-1\right) , \\
\left( 0,0,0,0,1,1\right) .
\end{array}
\end{equation}
The metric induced on $N$ is twice a four by four identity matrix. The
square root of its determinant is  four. Hence, the number of fixed
tori in that sector is 16/4 = 4.

In order to compute the Euler number we have to determine the points
which are left invariant under the action of two different non trivial
elements of ${\mathbb Z}_2 \times {\mathbb Z}_2$. For any such
combination we find that the third and fourth entry of the fixed point
vector have to vanish, whereas the first two and last two entries can
be any of the following four choices
\begin{equation}
\left(0,0\right) ,\,\,\, \left( 1,0\right) ,\,\,\, \left( \frac{1}{2},
\pm\frac{1}{2}\right) .
\end{equation}
Hence, we get
\begin{equation}
\chi_{\theta_1, \theta_2} = \chi_{\theta_1 , \theta_1 \theta_2} =
\chi_{ \theta_2, \theta_1 \theta_2} = 16,
\end{equation}
Note, that this time e.g.\ $\chi_{\theta_1, \theta_2}$ is not four
times the number of tori fixed under $\theta_1$. For the Euler number
we obtain
\begin{equation}
\chi =  24 ,
\end{equation}
which differs from twice the number of fixed tori.

\subsection{SO(6)$^2$ Lattice - Second Example} \label{sec:so6sec}

\vspace{.3in}

Now, we consider a version of the previous compactification
lattice which exhibits invariance under permuting the $x^1$-$x^2$,
$x^3$-$x^4$ and $x^5$-$x^6$ plane, i.e.\ we choose
\begin{eqnarray}
e_1 & = & \left( 1 ,0, -1 , 0,0,0\right) , \nonumber \\
e_2 & = & \left( 0, 0, 1, 0,-1,0\right) , \nonumber \\
e_3 & = & \left( 0 ,0, 1, 0,1,0\right) , \nonumber \\
e_4 & = & \left( 0, 1, 0, -1,0,0\right) , \nonumber \\
e_5 & = & \left( 0,0,0,1,0,-1\right) , \nonumber \\
e_6 & = & \left( 0,0,0,1,0,1\right) 
\label{eq:secso6lat}
\end{eqnarray}
as a basis for the SO(6)$^2$ lattice. For lattice vectors in the first
(second) SO(6) factor only odd (even) components can be non vanishing.

For that compactification one finds the following four inequivalent
fixed tori under the action of $\theta_2$ (\ref{eq:orbiact2}):
\begin{eqnarray}
\left\{ \left( x , y ,
  0,0,0,0\right) 
  \left|\, x,y \in {\mathbb R}^2/ 
\Lambda^2 \right. \right\} , & & \mbox{1 fixed torus},\label{eq:fix1t}\\
\left\{ \left( x , y ,
  \frac{1}{2},0,\frac{1}{2},0\right) 
  \left|\, x,y \in {\mathbb R}^2/ 
\Lambda^2 \right. \right\} , & & \mbox{1 fixed torus},\label{eq:fix2t} \\
\left\{ \left( x , y ,
  0,\frac{1}{2}, 0, \frac{1}{2}\right)
  \left|\, x,y \in {\mathbb R}^2/ 
\Lambda^2 \right. \right\} , & & \mbox{1 fixed
  torus}, \label{eq:fix3t}\\ 
\left\{ \left( x , y ,
  \frac{1}{2},\frac{1}{2},\frac{1}{2},\frac{1}{2}\right)
  \left|\, x,y \in {\mathbb R}^2/ 
\Lambda^2 \right. \right\} , & & \mbox{1 fixed
  torus}.\label{eq:fix4t}
\end{eqnarray}
For $\theta_1$ and $\theta_1 \theta_2$ one finds also four fixed tori
each in a completely analogous way.
So, altogether there are 12 fixed tori. Now, a single projection with
a non-trivial element of ${\mathbb Z}_2 \times {\mathbb Z}_2$ projects
out moduli needed for deforming the lattice into the SO(12)
lattice. The orbifolds are disconnected.

The computation using the Lefschetz fixed point
theorem (\ref{eq:lefschetz}) is similar to 
the last case in the previous example.
A basis for the $\left( 1 -
\theta_2\right) \Lambda $ lattice is
\begin{equation}
\left( 1 - \theta_2\right) \Lambda: \,\,\, 
\begin{array}{c} \left(0,0,2,0,0,0\right), \\
\left( 0,0,0,2,0,0\right) , \\
\left( 0,0,0,0,2,0\right) , \\
\left( 0,0,0,0,0,2\right) ,
\end{array}
\end{equation}
where the first basis vector originates e.g.\ from the SO(6)$^2$
lattice vector $(1,0,1,0,0,0)$. The volume of the fundamental cell is
16. The lattice normal to the invariant
lattice has the following basis:
\begin{equation}
N: \,\,\, 
\begin{array}{c} \left(0,0,1,0,-1,0\right), \\ 
\left( 0,0,1,0,1,0\right) , \\
\left( 0,0,0,1,0,-1\right) , \\
\left( 0,0,0,1,0,1\right) .
\end{array}
\end{equation}
The volume of the fundamental cell of $N$ is four. Hence, the number of
fixed tori under $\theta_1$ is 16/4 = 4. The computation for
$\theta_1$ and $\theta_1 \theta_2$ is analogous and yields four fixed
tori for each of them. 

The points which are fixed under any two non-trivial elements of
${\mathbb Z}_2 \times {\mathbb Z}_2$ can be written as
\begin{equation}
\left( a_1, b_1, a_2, b_2, a_3 , b_3 \right) ,
\end{equation}
where $\left( a_1, a_2 , a_3 \right)$ and $\left( b_1 , b_2 ,
b_3\right)$ can be any of the following four triplets
\begin{equation}
\left( 0,0,0\right) , \,\,\, \left( 1,0,0\right) , \,\,\, \left(
\frac{1}{2}, \frac{1}{2}, \pm \frac{1}{2}\right) .
\end{equation}
This gives 16 such points which is four times the number of tori fixed
under a single non-trivial ${\mathbb Z}_2 \times {\mathbb Z}_2$ element.
The Euler number is
\begin{equation}
\chi = 24,
\end{equation}
which is twice the number of all fixed tori.

\subsection{SU(3)$^3$ Lattice}\label{sec:su3}

Here, we give a second example for which the total number of fixed
tori differs from half the Euler number. The basis for the
compactification lattice is given by simple 
roots of SU(3)$^3$
\begin{align}
e_1 & =  \left( \sqrt{2},0,0,0,0,0\right) , \notag \displaybreak[2]\\
e_2 & =  \left( -\frac{1}{\sqrt{2}},0,0,
\sqrt{\frac{3}{2}},0,0\right) , \notag \displaybreak[2]\\
e_3 & =  \left( 0, \sqrt{2},0,0,0,0\right) , \notag \displaybreak[2]\\
e_4 & = \left( 0, -\frac{1}{\sqrt{2}},0,0, \sqrt{\frac{3}{2}}, 0
\right) , \notag\displaybreak[2] \\
e_5 & = \left( 0,0, \sqrt{2}, 0,0,0\right) , \notag\displaybreak[2] \\
e_6 & = \left( 0,0,-\frac{1}{\sqrt{2}},0,0,\sqrt{\frac{3}{2}}
\right) . \label{eq:su3qbasis}
\end{align}

For the above SU(3)$^3$ lattice one finds four fixed tori per
${\mathbb Z}_2 \times {\mathbb Z}_2$ twist. For instance, the
$\theta_2$ fixed tori are given by
\begin{eqnarray}
\left\{ \left( x , y ,
  0,0,0,0\right) 
  \left|\, x,y \in {\mathbb R}^2/ 
\tilde{\Lambda}^2/\sqrt{2} \right. \right\} , & & \mbox{1 fixed
  torus},
\label{eq:su3f1}\\
\left\{ \left( x , y ,
  \frac{1}{\sqrt{2}},0,0,0\right) 
  \left|\, x,y \in {\mathbb R}^2/ 
\tilde{\Lambda}^2/\sqrt{2} \right. \right\} , & & \mbox{1 fixed
  torus}, 
\label{eq:su3f2}\\
\left\{ \left( x , y ,
  -\frac{1}{2\sqrt{2}}, 0,0, \frac{1}{2}\sqrt{\frac{3}{2}}\right)
  \left|\, x,y \in {\mathbb R}^2/ 
\tilde{\Lambda}^2/\sqrt{2} \right. \right\} , & & \mbox{1 fixed
  torus}, \label{eq:su3f3}\\ 
\left\{ \left( x , y ,
  \frac{1}{2\sqrt{2}}, 0,0, \frac{1}{2}\sqrt{\frac{3}{2}}\right)
  \left|\, x,y \in {\mathbb R}^2/ 
\tilde{\Lambda}^2/\sqrt{2} \right. \right\} , & & \mbox{1 fixed
  torus} .\label{eq:su3f4}
\end{eqnarray}
In a similar fashion one finds four fixed tori also for the other two
non-trivial elements of ${\mathbb Z}_2 \times {\mathbb Z}_2$ yielding
to a total number of twelve fixed tori.

In the following we compute the number of tori invariant under the
action of $\theta_2$ using the Lefschetz fixed point theorem
(\ref{eq:lefschetz}). The lattice $\left( 1 - \theta_2\right) \Lambda$
is generated by
\begin{equation}
\left( 1 - \theta_2\right) \Lambda: \,\,\, 
\begin{array}{c} \left(0,0,2\sqrt{2},0,0,0\right), \\
\left( 0,0,0,2\sqrt{\frac{3}{2}},0,0\right) , \\
\left( 0,0,0,0,2\sqrt{\frac{3}{2}},0\right) , \\
\left( 0,0,-\sqrt{2},0,0,2\sqrt{\frac{3}{2}}\right) .
\end{array}
\end{equation}
Hence, the induced metric is
$$\left( \begin{array}{r r r r}
8 & 0 & 0 & -4 \\
0 & 6 & 0 & 0  \\
0 & 0 & 6 & 0  \\
-4 & 0 & 0 & 8 \end{array}\right) , $$
with determinant 36$\cdot$48.

The lattice $N$ normal to the invariant lattice is
\begin{equation}
N: \,\,\, 
\begin{array}{c} \left(0,0,\sqrt{2},0,0,0\right), \\
\left( 0,0,0,2\sqrt{\frac{3}{2}},0,0\right) , \\
\left( 0,0,0,0,2\sqrt{\frac{3}{2}},0\right) , \\
\left( 0,0,-\frac{\sqrt{2}}{2},0,0,\sqrt{\frac{3}{2}}\right) ,
\end{array}
\end{equation}
with  the induced metric
$$\left( \begin{array}{r r r r}
2 & 0 & 0 & -1 \\
0 & 6 & 0 & 0  \\
0 & 0 & 6 & 0  \\
-1 & 0 & 0 & 2 \end{array}\right) , $$
whose determinant is 36$\cdot$3. The number of fixed tori is the
square root of the ratio of the two determinants which is four in the
considered case. Taking into account also the other ${\mathbb Z}_2
\times {\mathbb Z}_2$ twists we find 12 fixed tori altogether.

There are eight points on the lattice which are invariant under
$\theta_1$ and $\theta_2$. They are of the form that the last three
entries of their position vector vanish whereas each of the first three
entries can be either zero or $\sqrt{2}/4$. Doing similar countings
also for the other combinations of ${\mathbb Z}_2 \times {\mathbb
  Z}_2$ elements leads finally to the Euler number,
\begin{equation}
\chi = 12,
\end{equation}
which differs from twice the number of fixed tori.

\section{Standard Embedding} \label{sec:SE}

Now, we consider heterotic E$_8 \times$E$_8$ theory compactified on
the previously discussed orbifolds. The massless spectrum in four
dimensions does not contain winding or momentum modes (away from the
selfdual point). To a large
extent, the calculation of the spectrum is independent of the
compactification lattice. The lattice enters via its fixed point
structure as will be seen shortly.

In this section, we focus on the standard embedding of the orbifold
action into the gauge group. In that case the spin connection is
related to the gauge connection. This implies that the index of the
Dirac operator is fixed by the Euler number which therefore equals
twice the number of generations \cite{Dixon:1985jw,Dixon:1986jc}. 

First, let us summarise what is known from previous
calculations \cite{Font:1988mk,nfw}. The untwisted sector
provides geometric moduli, vector multiplets containing the gauge fields
for an unbroken
\begin{equation}\label{eq:4dgaugegroup}
\mbox{E}_6\times \mbox{U(1)}^2 \times \mbox{E}^\prime _8 
\end{equation}
symmetry. Further, there are three chiral multiplets transforming in
the {\bf 27} and three chiral multiplets in the {\boldmath{$\overline{
    27}$}} of E$_6$ (we suppress U(1) charges). In addition there are
six singlets. 

Each non-trivial element of ${\mathbb Z}_2 \times {\mathbb Z}_2$ gives
rise to twisted sector states localised to the corresponding fixed
tori. Fields appearing in the bulk of a fixed torus are even under
the ${\mathbb Z}_2$ element leaving that torus invariant. The
remaining ${\mathbb Z}_2 \times {\mathbb Z}_2$ elements relate field
values at a point to values at an image point and act as projections
if the argument of the field is a fixed point.

The gauge group in the bulk of the fixed tori is
$$ \mbox{E}_7 \times \mbox{SU(2)} \times \mbox{E}_8 ^\prime . $$
The twisted sector gives rise to bulk matter (hypermultiplets)
transforming in the ({\bf 56},{\bf 1}) plus ({\bf 1},{\bf 2}) of E$_7
\times$SU(2) (w.r.t.\ the hidden E$_8 ^\prime$ all matter fields are
singlets).      

The remaining ${\mathbb Z}_2$ projection reduces the bulk gauge group
to the group unbroken in four dimensions (\ref{eq:4dgaugegroup}). The
projections on the twisted matter depend on the compactification
lattice\footnote{In technical terms: The centraliser
  \cite{Dixon:1985jw} of a space group 
  element depends on the compactification lattice.}. Let us first
state what happens in the well studied case that 
the underlying $T^6$ factorises into three $T^2$ factors
\cite{Font:1988mk,nfw}. In that case the hypermultiplet in
the ({\bf 56},{\bf 1}) plus ({\bf 1},{\bf 2}) is projected to a
chiral multiplet in the {\boldmath{$\overline{27}$}} of E$_6$ and five
more chiral 
multiplets which are E$_6$ singlets. Since this happens for each fixed
torus, the number of generations is equal to the number of fixed tori,
which is 48 in the considered case.

On the other hand, the net number of generations follows from an index
theorem to be half the Euler number. So, the statement that the number
of generations equals the number of fixed tori cannot be true in
general (not even in ${\mathbb Z}_2 \times {\mathbb Z}_2$ orbifolds)
since we have seen that, in certain cases, the two numbers do not agree.
(For the example considered above the Euler number is 96 and the
result is consistent.)

In the following we will have a closer look at the projections of bulk
matter of a $\theta_i$ fixed torus, focusing on matter coming from the
$\theta_i$ twisted sector. First, we study the example from section
\ref{sec:so12} (compactification lattice equals SO(12) root lattice)
for which the Euler number is twice the number of fixed 
tori.  

The $\theta_2$ twisted sector provides eight hypermultiplets in the 
({\bf 56},{\bf 1}) + ({\bf 1},{\bf 2}) of E$_7 \times$SU(2), one on
each of the 8 fixed tori
\begin{eqnarray}
\left\{ \left( x , y ,
  0,0,0,0\right) 
  \left|\, x,y \in {\mathbb R}^2/ 
\Lambda^2 \right. \right\} , & & \mbox{1 fixed torus},\\
\left\{ \left( x , y ,
  \underline{\frac{1}{2},\frac{1}{2},0,0}\right) 
  \left|\, x,y \in {\mathbb R}^2/ 
\Lambda^2 \right. \right\} , & & \mbox{6 fixed tori}, \\
\left\{ \left( x , y ,
  \frac{1}{2},\frac{1}{2},\frac{1}{2},\frac{1}{2}\right)
  \left|\, x,y \in {\mathbb R}^2 
\Lambda^2 \right. \right\} , & & \mbox{1 fixed
  torus}. 
\end{eqnarray}
Now, we discuss at which points $\theta_1$ invariance provides
projection conditions on these hypermultiplets. 
First, consider the fixed tori of the form
\begin{eqnarray}
\left\{ \left( x , y ,
  0,0,0,0\right) 
  \left|\, x,y \in {\mathbb R}^2/ 
\Lambda^2 \right. \right\} ,\nonumber \\
\left\{ \left( x , y ,
  \frac{1}{2},\frac{1}{2},0,0\right) 
  \left|\, x,y \in {\mathbb R}^2/ 
\Lambda^2 \right. \right\} , \nonumber \\
\left\{ \left( x , y ,0,0,
  \frac{1}{2},\frac{1}{2}\right) 
  \left|\, x,y \in {\mathbb R}^2/ 
\Lambda^2 \right. \right\} , \nonumber \\
\left\{ \left( x , y ,
  \frac{1}{2},\frac{1}{2},\frac{1}{2},\frac{1}{2}\right)
  \left|\, x,y \in {\mathbb R}^2/ 
\Lambda^2 \right. \right\} . \label{eq:fixed1stset}
\end{eqnarray}
Taking into account SO(12) lattice shifts $\theta_1$ maps a point on
one of these tori to a point on the same torus with $(x,y) \to
-(x,y)$. Projection conditions thus arise at the points for which
$$(x,y) = - (x,y) \,\,\,\mbox{mod } \Lambda^2 , $$ 
which is satisfied for the following four points
\begin{equation}
(0,0) , \,\,\, \left(\frac{1}{2}, -\frac{1}{2}\right) ,\,\,\,
  \left(\frac{1}{2}, 
  \frac{1}{2}\right) ,\,\,\, (1,0) . 
\end{equation}
Imposing invariance under $\theta_1$ at these points reduces the
hypermultiplets to chiral ones in {\boldmath{$\overline{27}$}} of
E$_6$ and 
singlets. So, each of the four tori in (\ref{eq:fixed1stset}) gives
rise to one generation.

Points on the remaining four fixed tori transform as
\begin{equation}
\left( x, y, \underline{\frac{1}{2}, 0}, \underline{\frac{1}{2}, 0}
  \right) \stackrel{\theta_1}{\longrightarrow} \left( -x, -y,
  \underline{-\frac{1}{2}, 0}, \underline{\frac{1}{2}, 0} \right)
\equiv \left( -x +1 , -y, \underline{\frac{1}{2}, 0},
  \underline{\frac{1}{2}, 0} \right),
\end{equation}
where in the last step equivalence under shifts by the SO(12) lattice
vectors 
\begin{equation}\label{eq:crossshifts}
\left( 1 ,0,0, \underline{1,0}, 0,0\right)
\end{equation}
has been
employed. Thus $\theta_1$ maps a point on a fixed torus to itself if
$(x,y)$ is at one of 
the following four points\footnote{Considering an effective six
  dimensional orbifold GUT it looks a bit strange that we have two
  sets with four fixed points in each set. However, such an effective
  six dimensional orbifold GUT is obtained by taking the volume of the
  fixed torus to be much larger than the volume of the remaining
  compact space (by tuning metric moduli). Then shifts of the form
  (\ref{eq:crossshifts}) appear as an effective symmetry $x \to x+1$
  resulting in a compactification torus whose lattice is generated by
  $(1,0)$ and $(0,1)$. On that smaller torus some fixed points become
  indistinguishable, and the effective six dimensional orbifold GUT is
  compactified on $T^2/{\mathbb Z}_2$.}
\begin{equation}
\left( \frac{1}{2} , 0 \right) ,\,\,\, \left(\frac{3}{2},
\frac{1}{2}\right) ,\,\,\, \left( \frac{3}{2}, -\frac{1}{2}\right) ,
\,\,\, \left(\frac{3}{2},0\right) .
\end{equation}
So, also each hypermultiplet on these remaining fixed tori gives rise
to a chiral multiplet transforming in the {\boldmath{$\overline{27}$}}
of E$_6$ and 
singlets. 

Repeating our discussion for the remaining two twist sectors, we find
that the number of generations is equal to the number of fixed tori,
i.e.\ 24. Since in the considered case the Euler number is 48, our
result is consistent with the index theorem.  

As a next case we study the example of section \ref{sec:so6-1}. In
this case the Euler number differs from twice the number of fixed
tori. First, let us look at the $\theta_2$ twisted sector. Again,
there is a hypermultiplet in ({\bf 56},{\bf 1}) + ({\bf 1},{\bf 2})
on each of the $\theta_2$ fixed tori (\ref{eq:so6fixf})--
(\ref{eq:minusnomat}). For points on the first two fixed tori
(\ref{eq:so6fixf}) and (\ref{eq:sector}) the situation is the same as
in the previous example in that these points are mapped onto points in
the same torus with four points being invariant. Hence, these two
fixed tori result in two chiral multiplets in the
{\boldmath{$\overline{27}$}} of E$_6$ plus singlets. 
The same happens for fixed tori of the form
\begin{eqnarray}
\left\{ \left( x , y ,
  0,0,\frac{1}{2},\frac{1}{2}\right)
  \left|\, x,y \in {\mathbb R}^2/ 
\Lambda^2 \right. \right\} , \\ 
\left\{ \left( x , y ,
  0,0,\frac{1}{2},-\frac{1}{2}\right)
  \left|\, x,y \in {\mathbb R}^2/ 
\Lambda^2 \right. \right\} ,\\
\end{eqnarray}
giving rise to two more chiral multiplets in the
{\boldmath{$\overline{27}$}} and singlets. 
For the remaining four fixed tori
\begin{eqnarray}
\left\{ \left( x , y ,
  0,\frac{1}{2},\underline{0,\frac{1}{2}}\right)
  \left|\, x,y \in {\mathbb R}^2/ 
\Lambda^2 \right. \right\} ,\label{eq:nonch1} \\ 
\left\{ \left( x , y ,
  0,\frac{1}{2},\underline{0,-\frac{1}{2}}\right)
  \left|\, x,y \in {\mathbb R}^2/ 
\Lambda^2 \right. \right\} ,\label{eq:nonch2}
\end{eqnarray}
the situation is different. 
Points in tori of the form (\ref{eq:nonch1}) are mapped to points on
tori of the form (\ref{eq:nonch2}). This means that projection conditions
identify hypermultiplets, leading to two massless hypermultiplets
in four dimensions. In $N=1$ language, one hypermultiplet decomposes
into a chiral multiplet in the {\bf 27}, a chiral
multiplet in the
$\overline{\bf 27}$ and singlets. Hence, we get two generations and
two anti-generations from the twisted states on  (\ref{eq:nonch1})
and (\ref{eq:nonch2}). By employing the symmetry between the two
SO(6) factors in the compactification lattice, we deduce that the
$\theta_1$ twisted sector also gives six chiral multiplets in the
{\bf 27} and two chiral multiplets in the $\overline{\bf 27}$, with
a net number of four generations.

$\theta_1\theta_2$ fixed tori are listed in (\ref{eq:onetwof}) --
(\ref{eq:onetwof}). Points on each of these tori are mapped onto
points on the same torus. Hence, the number of generations coming from
the $\theta_1\theta_2$ twisted sector is four.

In summary, for the example of section \ref{sec:so6-1} we obtain 18
generations and six anti-generations, from twisted sectors. This
relates to the 
topological data as follows: The net number of generations equals half
the Euler number and the surplus in the number of fixed tori provides
pairs of generations and anti-generations.

In the example of section \ref{sec:so6sec}, points on a fixed torus
are always mapped on points of the same torus. The number of
generations is equal to the number of fixed tori, which is the same as
half the Euler number.
  
Finally, we discuss the example of section \ref{sec:su3}. The tori
fixed under the action of $\theta_2$ are listed in (\ref{eq:su3f1}) --
(\ref{eq:su3f4}). Points on the first two fixed tori, (\ref{eq:su3f1})
and (\ref{eq:su3f2}), are mapped on points of the same torus by
$\theta_1$. Hence, these two tori give rise to two generations in four
dimensions. On the other hand, $\theta_1$ maps points on the fixed
torus (\ref{eq:su3f3}) to points on (\ref{eq:su3f4}). These two tori
result in a generation and an anti-generation. Repeating the
discussion for the other orbifold twist, we find from twisted sectors
nine generations and three anti-generations. The net number six
coincides with half the Euler number.  

We summarise our examples in table \ref{tab:summary}, where we also
added the untwisted sector providing three generations and three
anti-generations. 
\begin{table}
\begin{tabular}{| l | c | c |}
\hline 
Lattice from Section: & (generations , anti-generations) & net number
of generations \\
\hline \hline
\ref{sec:so12} \hfill (SO(12))& (27,3) & 24 \\
\hline
\ref{sec:so6-1} \hfill (SO(6)$^2$-A)& (19,7) & 12 \\
\hline
\ref{sec:so6sec} \hfill (SO(6)$^2$-B)& (15,3) & 12 \\
\hline 
\ref{sec:su3} \hfill (SU(3)$^3$)& (12,6) & 6 \\
\hline
\end{tabular}
\caption{Number of generations for standard embedding.}
\label{tab:summary}
\end{table} 
In addition, the number of fixed tori and the Euler number is listed
in table \ref{tab:top}.

\begin{table}
\begin{center}
\begin{tabular}{| l | c | c |}
\hline 
Lattice from Section: & fixed tori & Euler number \\
\hline \hline
\ref{sec:so12} \hfill (SO(12))& 24 & 48 \\
\hline
\ref{sec:so6-1} \hfill (SO(6)$^2$-A)& 20  & 24 \\
\hline
\ref{sec:so6sec} \hfill (SO(6)$^2$-B)& 12 & 24 \\
\hline 
\ref{sec:su3} \hfill (SU(3)$^3$)& 12 & 12 \\
\hline
\end{tabular}
\caption{Number of fixed tori and Euler number.}
\label{tab:top}
\end{center}
\end{table} 
These results can be confirmed by deriving projections from modular
invariance as was done in \cite{Ibanez:1987pj} and extended in
\cite{Erler:1992ki} for the case of fixed tori and non factorisable
lattices. The strategy is to split the partition function into various
pieces corresponding to different twist sectors and different
insertions of orbifold group elements into the trace. Such splitting
mixes under modular transformations and hence modular
invariance fixes numerical factors in front of some terms. The number
of generations and anti-generations can be deduced by identifying
contributions from chiral and anti-chiral fermions to the trace.   

A subtlety for non factorisable compactifications is the existence of
contributions in which winding and momentum sums are not over mutually
dual lattices. For instance, in the trace over untwisted
states with a $\theta_i$ insertion only windings on an invariant
sublattice contribute, whereas momenta are constrained to an invariant
sublattice of the dual compactification lattice. The latter is the
dual of a $\theta_i$ projected lattice which differs from the
$\theta_i$ invariant lattice in the non factorisable case. Modular
invariance relates that contribution to the $\theta_i$ twisted
sector. There, windings are constrained to the $\theta_i$ projected
lattice, since strings have to close only up to a $\theta_i$
identification. (The invariant lattice is a sublattice of the
projected lattice.) Momenta, on the other hand, take values on the dual
of the $\theta_i$ invariant lattice which spans the $\theta_i$ fixed
torus. 

Thus we observe that the two contributions discussed above are related
by interchanging winding and momenta and dualising the involved
lattices. This can be achieved by Poisson 
resummation and hence corresponds to modular
transformations. Numerical factors equal to ratios of the volumes of
invariant and projected lattices appear, and this explains the
lattice dependence in the number of families and anti-families. (We
have carried out a detailed analysis of all cases discussed in this
paper and found the correct numerical factors.) 

Another subtlety which we did not mention so far is the ambiguity in the
choice of sign for some of the different contributions to the
partition function. This corresponds to the inclusion of discrete
torsion \cite{Vafa:1986wx,Font:1988mk,Vafa:1994rv}. In the ${\mathbb
  Z}_2 \times {\mathbb Z}_2$ case the terms in question are traces
over $\theta_i$ twisted sectors with $\theta_j$ ($i \not= j$)
insertions. In these traces chiral and anti-chiral fermions contribute
with opposite signs. Hence, switching on and off discrete torsion just
swaps the number of families with the number of anti-families (see
\cite{Font:1988mk}). As discussed in
\cite{Vafa:1994rv}, for the factorisable case, this corresponds to
replacing the Hodge diamond by its mirror. Our discussion suggests
that the cases with and without discrete torsion are related by mirror
symmetry also if the underlying $T^6$ does not factorise.

Finally, a comment about non-standard embeddings is in order. Since
we discussed each twisted sector separately it should be
straightforward to modify our considerations for the other consistent
choices of the orbifold embedding into the gauge group (listed in
\cite{Forste:2005gc}).  

\section{A Three Generation Model}

In this section we present a concrete three generation model
with an unbroken SO(10) gauge group. The orbifold will be standard
embedded and from the previous section it is clear that we need to add
Wilson lines \cite{Ibanez:1986tp}. The
number of models with Wilson lines is large and, in order to ease our
task, we seek some intuition. 

First, we will focus on models with
SO(10) gauge group. This simplifies the counting of generations since
one generation corresponds to a 16 dimensional representation. From a
phenomenological perspective, the construction of SO(10) models can be
viewed as a step towards obtaining realistic models. This is suggested
by the free fermionic models discussed in section two, as well as the
geometric picture \cite{nfw,Nilles:2004ej}. 

Secondly, we borrow some more intuition from the free fermionic
construction by looking for models where each twisted sector provides
one generation. For ${\mathbb Z}_3$ orbifolds, models where the number
of generations is associated to the number of complex extra dimensions
\cite{Ibanez:1986tp} as well as to multiplicities of twisted sectors
\cite{Kim:1992en} have been constructed. In the ${\mathbb Z}_2 \times
{\mathbb Z}_2$ case, however, most of the known three generation
models \cite{nfw} do not exhibit such a structure. The untwisted
spectrum is 
always non chiral and trying to associate the number of generations to
the complex number of extra dimensions is hopeless. 

So, it remains to look for models where the three twisted sectors
contribute one generation each. As we will argue now, the fact whether
or not the underlying $T^6$ factorises into the product of three $T^2$ 
is essential in that context. In order to break the E$_6$ gauge
symmetry down to SO(10) one needs to turn on a Wilson line removing
the 16 dimensional spinors of SO(10) from the adjoint of E$_6$. In the
case of a 
factorisable $T^6$, the cycle along which this Wilson line is turned
on will be invariant under one of the three non trivial ${\mathbb Z}_2
\times {\mathbb Z}_2$ elements. Within the corresponding twisted
sector, the Wilson line will impose a projection removing the 16
dimensional representations of SO(10). Thus that sector cannot
contribute to the number of generations.\footnote{Our argument can be
evaded by a more baroque embedding of SO(10) into E$_6$. An example is
reported in \cite{patrick}. We thank Patrick Vaudrevange for drawing
our attention to that possibility.}   

For non factorisable tori, on the other hand, there exist cycles which
are invariant under none of the three non trivial ${\mathbb Z}_2
\times {\mathbb Z}_2$ elements. Turning on the Wilson line breaking
E$_6$ to SO(10) on such a cycle does not necessarily remove {\bf 16}
representations from one of the twisted sectors.  

In the explicit calculation one has to determine the contribution from
each fixed torus to the spectrum. So, as a final simplification, we
focus on models with a minimal number of fixed tori.

Therefore, we consider the
SO(6)$^2$ lattice of section \ref{sec:so6sec} where we will indeed
find a three generation model with SO(10) gauge symmetry. Before
giving the details of that model, let us discuss consistency conditions
for the Wilson lines. For the present considerations it is useful to
view the Wilson line as a vacuum expectation value for an internal
gauge field component, $A_i$, where the index labels directions along
lattice vectors (\ref{eq:secso6lat}) (see also (\ref{lrmomenta}) for
the notation). For discrete Wilson lines the value of $A_i$ can differ
from its orbifold image only by E$_8 \times$E$_8$ root lattice vectors
\cite{Ibanez:1986tp}. In order to find the resulting consistency
condition we state the orbifold action on the vectors generating the
SO(6)$^2$ lattice  (\ref{eq:secso6lat})
\begin{equation}
\theta_1 :\,\, \begin{array}{l c l}
e_1 & \to & - e_1, \\
e_2 & \to & - e_3, \\
e_3 & \to & - e_2, \\
e_4 & \to & - e_4, \\
e_5 & \to & - e_6, \\
e_6 & \to & - e_5, 
\end{array}\hspace*{1in}
\theta_2 :\,\, \begin{array}{l c l}
e_1 & \to & e_1 + e_2 + e_3 , \\
e_2 & \to & - e_2 ,\\
e_3 & \to & - e_3 ,\\
e_4 & \to & e_4 + e_5 + e_6 , \\
e_5 & \to & - e_5 , \\
e_6 & \to & - e_6 .
\end{array} 
\end{equation}    
Hence we find the following consistency conditions
\begin{equation}
2 A_i , \,\,\, A_2 + A_3,\,\,\, A_5 + A_6  \in
\Lambda_{\mbox{\tiny E$_8\times$E$_8$}} ,\,\,\, i = 1, \ldots , 6
,
\end{equation}
where $\Lambda_{\mbox{\tiny E$_8\times$E$_8$}}$ denotes the E$_8
\times$E$_8$ root lattice. The other condition which has to be
satisfied comes from modular invariance (see e.g.\
\cite{Ibanez:1987pj}). 

Adopting the notation of e.g.\ \cite{nfw} we characterise the standard
embedding by the shift vectors
\begin{equation}
V_1 = \left( \frac{1}{2}, -\frac{1}{2}, 0^6\right)\left( 0^8\right)
\,\,\, , \,\,\, 
V_2 = \left( 0, \frac{1}{2}, -\frac{1}{2}, 0^5\right) \left(
0^8\right) .
\end{equation}
Further, we turn on the following set of consistent Wilson lines
\begin{eqnarray}
A_1 & = & \left( 0^8\right) \left( 0^3, \frac{1}{2}, \frac{1}{2},
-\frac{1}{2}, -\frac{1}{2}, 0\right) , \\
A_2 = A_3 & = & \left( 0^7,1\right) \left( 1, 0^7\right) ,
\label{eq:remspin}\\ 
A_4 & = & \left( 0^8\right) \left( 0 , -\frac{1}{2}, -\frac{1}{2},
0,0, \frac{1}{2}, \frac{1}{2},0\right) , \\
A_5 = A_6 & = & \left( 0^8\right) \left( 0, \frac{1}{2}, \frac{1}{2},
-\frac{1}{2}, -\frac{1}{2}, 0^3 \right) .
\end{eqnarray}
For the computation of the twisted sectors one has to identify the
space group element leaving a given torus invariant. For instance, the
torus (\ref{eq:fix1t}) is fixed under the rotation by $\theta_2$,
whereas for the tori
(\ref{eq:fix2t}), (\ref{eq:fix3t}), (\ref{eq:fix4t}) the rotation by
$\theta_2$ has to be supplemented by a 
shift with $e_3$, $e_6$, $e_3 + e_6$, respectively. With these remarks
the calculation of the massless spectrum follows straightforward
modifications 
of standard techniques and we refrain from giving a detailed step by
step presentation. Instead, we just report the result. 

The metric and
$B$ field moduli are $g_{IJ}$ and $b_{IJ}$ with $\left( I ,J\right) =
\left( 1,2\right)$, $\left( 3,4\right)$ or $\left( 5,6\right)$, where
the indices $I,J$ label Cartesian coordinates. We choose these moduli
to be different from their selfdual values in order not to have to
report more massless states than necessary. 

The untwisted spectrum gives rise to an $N=1$ vector multiplet in the
adjoint of
\begin{equation}\label{eq:unbroken}
\mbox{SO(10)} \times \mbox{U(1)}^3 \times 
\mbox{SU(2)}^8 ,
\end{equation}
where the SO(10) and U(1) factors come from the first E$_8$ factor
whereas the SU(2) factors originate from the second (hidden) E$_8$
factor. The three U(1) factors are generated by the first three Cartan
operators of the first E$_8$: $H_1$, $H_2$, $H_3$. The remaining
Cartan generators of the first E$_8$, $H_3, \ldots , H_8$, form the
Cartan Subalgebra of SO(10). Denoting the Cartan generators of the
second E$_8$ by $H_9,
\ldots , H_{16}$ we arrange the eight SU(2) factors according to the
following order of their Cartan generators:
\begin{equation}
H_9\! +\! H_{16} , H_9\! -\! H_{16}, H_{10}\! +\! H_{11}, H_{10}\!
-\! H_{11}, 
H_{12}\! +\! H_{13},  H_{12}\! -\! H_{13},  H_{14}\! +\! H_{15},
H_{14}\! - \! H_{15} .
\end{equation}
Below we will characterise representations of the unbroken gauge group
(\ref{eq:unbroken}) by an ordered nonet containing the
dimensionality of the SO(10) representation followed by the
dimensionalities of the SU(2)$^8$ representation. A sequence of $n$
consecutive entries being {\bf 1} will be abbreviated by ${\bf
1}^n$. Further, a 
triple 
subscript will denote the U(1) charges.\footnote{For the impatient reader:
The important part of the result is that we find one 16 dimensional
representation of SO(10) in each twisted sector. These representations
are localised on fixed tori situated at the origin of the remaining
four directions.}

In the {\bf untwisted sector} one finds chiral multiplets transforming as
\begin{eqnarray}
\left({\bf 10};{\bf 1}^8\right)_{1,0,0}  + \left(
{\bf 1};{\bf 1}^8\right)_{0,1,1} + \left( {\bf 1} ; {\bf 1}^8\right)
_{0,1,-1} +\mbox{Charge 
Conjugate} , \label{eq:u1}\\ 
\left({\bf 10};{\bf 1}^8\right)_{0,1,0}  + \left(
{\bf 1};{\bf 1}^8\right)_{1,0,1} + \left( {\bf 1} ; {\bf 1}^8\right)
_{1,0,-1}  +\mbox{Charge 
Conjugate} ,\label{eq:u2}\\
\left({\bf 10};{\bf 1}^8\right)_{0,0,1}  + \left(
{\bf 1};{\bf 1}^8\right)_{1,1,0} + \left( {\bf 1} ; {\bf 1}^8\right)
_{1,-1,0}  +\mbox{Charge 
Conjugate} \label{eq:u3} ,    
\end{eqnarray}
where (\ref{eq:u1}), (\ref{eq:u2}), (\ref{eq:u3}) correspond to the
first, second and third complex plane, respectively.

Now, we turn to the {\bf \boldmath{$\theta_1$}
 twisted sector}.
The four $\theta_1$ fixed tori are
\begin{eqnarray}
\left\{ \left(
  0,0,0,0, x , y \right) 
  \left|\, x,y \in {\mathbb R}^2/ 
\Lambda^2 \right. \right\} ,\label{eq:fix1t1}\\
\left\{ \left(
  \frac{1}{2},0,\frac{1}{2},0, x , y \right) 
  \left|\, x,y \in {\mathbb R}^2/ 
\Lambda^2 \right. \right\} ,\label{eq:fix2t1} \\
\left\{ \left(
  0,\frac{1}{2}, 0, \frac{1}{2},  x , y \right)
  \left|\, x,y \in {\mathbb R}^2/ 
\Lambda^2 \right. \right\} , \label{eq:fix3t1}\\ 
\left\{ \left( 
  \frac{1}{2},\frac{1}{2},\frac{1}{2},\frac{1}{2}, x , y \right)
  \left|\, x,y \in {\mathbb R}^2/ 
\Lambda^2 \right. \right\} .\label{eq:fix4t1}
\end{eqnarray}
On each of these tori massless chiral multiplets are localised. Below,
we list the representations in which they transform:
\begin{itemize}
\item torus (\ref{eq:fix1t1}):
\begin{eqnarray}
& & \left( {\bf 16}; {\bf 1}^8\right)_{0,0,\frac{1}{2}} + \left( {\bf 10};
{\bf 1}^8\right) _{-\frac{1}{2}, -\frac{1}{2}, 0} + \left( {\bf
1}; {\bf 1}^8\right)_{\frac{1}{2}, \frac{1}{2}, 1} + \left( {\bf
1}; {\bf 1}^8\right)_{\frac{1}{2}, \frac{1}{2}, -1} +\nonumber \\ & &
2 \left( {\bf 
1}; {\bf 1}^8\right)_{\frac{1}{2}, -\frac{1}{2}, 0} + 2 \left( {\bf
1}; {\bf 1}^8\right)_{-\frac{1}{2}, \frac{1}{2}, 0} ,
\end{eqnarray}
\item torus (\ref{eq:fix2t1}) : 
\begin{eqnarray}
 & & \left( {\bf 1}; {\bf 2}, {\bf 1}, {\bf 2}, {\bf 1}^5\right)
_{-\frac{1}{2}, \frac{1}{2}, 0} + \left( {\bf 1}; {\bf 1}, {\bf 2},
{\bf 1}, {\bf 2}, {\bf 1}^4\right)_{-\frac{1}{2}, \frac{1}{2}, 0} + 
\nonumber \\ & &
\left( {\bf 1}; {\bf 1}^4, {\bf 2}, {\bf 1}, {\bf 2}, {\bf 1}\right)
_{\frac{1}{2}, -\frac{1}{2}, 0} + \left( {\bf 1}; {\bf 1}^5 , {\bf 2},
{\bf 1}, {\bf 2}\right)_{\frac{1}{2}, -\frac{1}{2}, 0} ,
\end{eqnarray}
\item torus (\ref{eq:fix3t1}):
\begin{eqnarray}
 & & \left( {\bf 1}; {\bf 2}, {\bf 1}^3, {\bf 2}, {\bf 1}\right)
_{\frac{1}{2}, -\frac{1}{2}, 0} + \left({\bf 1}; {\bf 1}^2, {\bf 2}, {\bf
1}^3, {\bf 2}, {\bf 1}\right)_{\frac{1}{2}, -\frac{1}{2},
0} +  
\nonumber \\ & &
\left( {\bf 1}; {\bf 1}, {\bf 2}, {\bf 1}^3, {\bf 2}\right)
_{-\frac{1}{2}, \frac{1}{2}, 0} + \left( {\bf 1}; {\bf 1}^3 , {\bf 2},
{\bf 1}^3, {\bf 2}\right)_{-\frac{1}{2}, \frac{1}{2}, 0} ,
\end{eqnarray}
\item torus (\ref{eq:fix4t1}):
\begin{eqnarray}
 & & \left( {\bf 1}; {\bf 2}, {\bf 1}^5, {\bf 2}, {\bf 1}\right)
_{\frac{1}{2}, -\frac{1}{2}, 0} + \left( {\bf 1}; {\bf 1}, {\bf 2},
{\bf 1}^5, {\bf 2} \right)_{\frac{1}{2}, -\frac{1}{2}, 0} + 
\nonumber \\ & &
\left( {\bf 1}; {\bf 1}^2, {\bf 2}, {\bf 1}, {\bf 2}, {\bf 1}^3\right)
_{-\frac{1}{2}, \frac{1}{2}, 0} + \left( {\bf 1}; {\bf 1}^3 , {\bf 2},
{\bf 1}, {\bf 2}, {\bf 1}^2\right)_{-\frac{1}{2}, \frac{1}{2}, 0} .
\end{eqnarray}
\end{itemize}
So, there is one generation on the torus (\ref{eq:fix1t1}).

In the {\bf \boldmath{$\theta_2$} twisted sector} one finds chiral
multiplets 
localised on the tori (\ref{eq:fix1t}) -- (\ref{eq:fix4t}). The
representations are
\begin{itemize}
\item torus (\ref{eq:fix1t}):
\begin{eqnarray}
& & \left( {\bf 16}; {\bf 1}^8\right)_{\frac{1}{2},0,0} + \left( {\bf 10};
{\bf 1}^8\right) _{0,-\frac{1}{2}, -\frac{1}{2}} + \left( {\bf
1}; {\bf 1}^8\right)_{1,\frac{1}{2}, \frac{1}{2}} + \left( {\bf
1}; {\bf 1}^8\right)_{-1,\frac{1}{2}, \frac{1}{2}} +\nonumber \\ & &
2 \left( {\bf 
1}; {\bf 1}^8\right)_{0,\frac{1}{2}, -\frac{1}{2}} + 2 \left( {\bf
1}; {\bf 1}^8\right)_{0, -\frac{1}{2}, \frac{1}{2}} ,
\end{eqnarray}
\item torus (\ref{eq:fix2t}) : 
\begin{eqnarray}
 & & \left( {\bf 1}; {\bf 2}, {\bf 2}, {\bf 1}^6\right)
_{0, -\frac{1}{2}, -\frac{1}{2}} + \left( {\bf 1}; {\bf 1}^2, {\bf 2},
{\bf 2}, {\bf 1}^4\right)_{0, -\frac{1}{2}, -\frac{1}{2}} + 
\nonumber \\ & &
\left( {\bf 1}; {\bf 1}^4, {\bf 2}, {\bf 2}, {\bf 1}^2\right)
_{0,\frac{1}{2}, \frac{1}{2}} + \left( {\bf 1}; {\bf 1}^6 , {\bf 2},
{\bf 2}\right)_{0,\frac{1}{2}, \frac{1}{2}} ,
\end{eqnarray}
\item torus (\ref{eq:fix3t}):
\begin{eqnarray}
 & & \left( {\bf 1}; {\bf 1}, {\bf 2}, {\bf 1}^5, {\bf 2}\right)
_{0, -\frac{1}{2}, \frac{1}{2}} + \left({\bf 1}; {\bf 1}^3, {\bf 2}, {\bf
1}, {\bf 2}, {\bf 1}^2\right)_{0,-\frac{1}{2}, \frac{1}{2}} +  
\nonumber \\ & &
\left( {\bf 1}; {\bf 2}, {\bf 1}^5, {\bf 2}, {\bf 1}\right)
_{0, \frac{1}{2}, -\frac{1}{2}} + \left( {\bf 1}; {\bf 1}^2 , {\bf 2},
{\bf 1}, {\bf 2}, {\bf 1}^2\right)_{0,\frac{1}{2}, -\frac{1}{2}} ,
\end{eqnarray}
\item torus (\ref{eq:fix4t}):
\begin{eqnarray}
 & & \left( {\bf 1}; {\bf 2}, {\bf 1}^6, {\bf 2}\right)
_{0,\frac{1}{2}, \frac{1}{2}} + \left( {\bf 1}; {\bf 1}, {\bf 2},
{\bf 1}^4, {\bf 2}, {\bf 1} \right)_{0,\frac{1}{2}, \frac{1}{2}} + 
\nonumber \\ & &
\left( {\bf 1}; {\bf 1}^2, {\bf 2}, {\bf 1}^2, {\bf 2}, {\bf 1}^2\right)
_{0,-\frac{1}{2}, \frac{1}{2}} + \left( {\bf 1}; {\bf 1}^3 , {\bf 2},
{\bf 2}, {\bf 1}^3\right)_{0,-\frac{1}{2}, \frac{1}{2}} .
\end{eqnarray}
\end{itemize}
There is one generation coming from the $\theta_2$ twisted sector
localised on the torus (\ref{eq:fix1t}).

{\bf \boldmath{$\theta_1\theta_2$} twisted sector} states are localised on the
following four tori:
\begin{eqnarray}
\left\{ \left(
  0,0, x , y, 0,0 \right) 
  \left|\, x,y \in {\mathbb R}^2/ 
\Lambda^2 \right. \right\} ,\label{eq:fix1t12}\\
\left\{ \left(
  \frac{1}{2},0, x , y,\frac{1}{2},0 \right) 
  \left|\, x,y \in {\mathbb R}^2/ 
\Lambda^2 \right. \right\} ,\label{eq:fix2t12} \\
\left\{ \left(
  0,\frac{1}{2}, x , y, 0, \frac{1}{2} \right)
  \left|\, x,y \in {\mathbb R}^2/ 
\Lambda^2 \right. \right\} , \label{eq:fix3t12}\\ 
\left\{ \left( 
  \frac{1}{2},\frac{1}{2}, x , y, \frac{1}{2},\frac{1}{2} \right)
  \left|\, x,y \in {\mathbb R}^2/ 
\Lambda^2 \right. \right\} .\label{eq:fix4t12}
\end{eqnarray}
The corresponding massless chiral multiplets transform as:
\begin{itemize}
\item torus (\ref{eq:fix1t12}):
\begin{eqnarray}
& & \left( {\bf 16}; {\bf 1}^8\right)_{0,\frac{1}{2},0} + \left( {\bf 10};
{\bf 1}^8\right) _{-\frac{1}{2},0, -\frac{1}{2}} + \left( {\bf
1}; {\bf 1}^8\right)_{\frac{1}{2},1, \frac{1}{2}} + \left( {\bf
1}; {\bf 1}^8\right)_{\frac{1}{2},-1, \frac{1}{2}} +\nonumber \\ & &
2 \left( {\bf 
1}; {\bf 1}^8\right)_{\frac{1}{2},0, -\frac{1}{2}} + 2 \left( {\bf
1}; {\bf 1}^8\right)_{-\frac{1}{2},0, \frac{1}{2}} ,
\end{eqnarray}
\item torus (\ref{eq:fix2t12}) : 
\begin{eqnarray}
 & & \left( {\bf 1}; {\bf 2}, {\bf 1}^2, {\bf 2}, {\bf 1}^4\right)
_{\frac{1}{2},0, \frac{1}{2}} + \left( {\bf 1}; {\bf 1}, {\bf 2},
{\bf 2}, {\bf 1}^5\right)_{\frac{1}{2},0, \frac{1}{2}} + 
\nonumber \\ & &
\left( {\bf 1}; {\bf 1}^4, {\bf 2}, {\bf 1}^2, {\bf 2}\right)
_{-\frac{1}{2},0, -\frac{1}{2}} + \left( {\bf 1}; {\bf 1}^5 , {\bf 2},
{\bf 2}, {\bf 1}\right)_{-\frac{1}{2},0, -\frac{1}{2}} ,
\end{eqnarray}
\item torus (\ref{eq:fix3t12}):
\begin{eqnarray}
 & & \left( {\bf 1}; {\bf 2}, {\bf 1}, {\bf 2}, {\bf 1}^5\right)
_{-\frac{1}{2},0, \frac{1}{2}} + \left({\bf 1}; {\bf 1}, {\bf 2}, {\bf
1}, {\bf 2}, {\bf 1}^4\right)_{\frac{1}{2},0,-\frac{1}{2}} +  
\nonumber \\ & &
\left( {\bf 1}; {\bf 1}^4, {\bf 2}, {\bf 1}, {\bf 2}, {\bf 1}\right)
_{-\frac{1}{2},0,\frac{1}{2}} + \left( {\bf 1}; {\bf 1}^5 , {\bf 2},
{\bf 1}, {\bf 2}, \right)_{\frac{1}{2},0, -\frac{1}{2}} ,
\end{eqnarray}
\item torus (\ref{eq:fix4t12}):
\begin{eqnarray}
 & & \left( {\bf 1}; {\bf 2}, {\bf 2}, {\bf 1}^6\right)
_{-\frac{1}{2},0,-\frac{1}{2}} + \left( {\bf 1}; {\bf 1}^2, {\bf 2},
{\bf 2}, {\bf 1}^4 \right)_{\frac{1}{2},0, \frac{1}{2}} + 
\nonumber \\ & &
\left( {\bf 1}; {\bf 1}^4, {\bf 2}, {\bf 2}, {\bf 1}^2\right)
_{\frac{1}{2},0, \frac{1}{2}} + \left( {\bf 1}; {\bf 1}^6 , {\bf 2},
{\bf 2}\right)_{-\frac{1}{2},0, -\frac{1}{2}} .
\end{eqnarray}
\end{itemize}
There is one generation localised on the fixed torus
(\ref{eq:fix1t12}).

The important part of the above spectrum is that there is one
generation from each twisted sector, yielding the observed value of
three generations in total. We have listed also all the other details
of the spectrum, because these enable one to perform non trivial
consistency checks. In the case at hand, all factors in the unbroken
gauge group are anomaly free. For the three U(1) factors this is a non
trivial statement and we have checked that the corresponding 36
triangle diagrams indeed vanish.

To conclude, let us summarise how the construction worked. By the
standard embedded orbifold one of the E$_8$ factors is broken to
E$_6$. The Wilson line (\ref{eq:remspin}) is responsible for breaking
E$_6$ down to SO(10), whereas all the remaining Wilson lines remove
{\bf 16} or $\overline{\bf 16}$ representations from all but one of
the four fixed tori in each twisted sector. Since we did not want to
break SO(10) further, we have chosen them such that they break only the
hidden sector gauge group. It is conceivable that, alternatively, one
can choose some of the additional Wilson lines such that SO(10) is
broken further to its Standard Model subgroup and, simultaneously, keep
the three 16 dimensional representations providing the Standard Model
matter plus three right handed neutrinos. This construction is to be
carried out in the near future \cite{progress}.

\section{Conclusions and Outlook}

The present paper started by summarising the free fermionic
constructions of 
four dimensional heterotic string theories.
We argued that the impressive success in reproducing essential
features of real particle physics should encourage us to seek similar
features in ${\mathbb Z}_2 \times {\mathbb Z}_2$ orbifold
compactifications of heterotic string theory. In particular, free
fermionic constructions suggest that the underlying six-dimensional
compactification lattice should be the SO(12) root lattice. 

Motivated by the above observations, we studied the geometric
properties of  ${\mathbb Z}_2 \times {\mathbb Z}_2$ orbifolds of
several non factorisable six-tori. In great 
detail, we discussed how to compute the number of fixed tori and the
Euler number. Subsequently, we related these data to the particle
spectrum, if the orbifold is standard embedded into the gauge group.
Without Wilson lines there are no three generation models.

By introducing discrete Wilson lines \cite{Ibanez:1986tp}, one can
break the gauge group and, simultaneously, reduce the number of
generations. We observed that for ${\mathbb Z}_2 \times {\mathbb
Z}_2$ orbifolds on non factorisable tori one can easily find models where
each of the three twisted sectors contributes exactly one generation,
whereas this is difficult in the case that the underlying $T^6$
factorises into a product of three $T^2$. This supports the conjecture
that free fermionic constructions are related to ${\mathbb Z}_2 \times
{\mathbb Z}_2$ orbifolds of non factorisable tori since also, in free
fermionic models, cases with the same distribution of the three
families exist.  
We illustrated the discussion by the explicit presentation of a
${\mathbb Z}_2 \times {\mathbb Z}_2$ orbifold of $T^6$ whose
compactification lattice is a root lattice of SO(6)$^2$.  
 
There are several directions to be explored in future work. To make
explicitly contact to the quasi--realistic
free fermionic models \cite{fsu5,fny,alr,euslm,nahe}, one should consider
the SO(12) root lattice as a compactification lattice. Realising these
models in the orbifold language would not only improve our
understanding of the correspondence between the two schemes, but also
provide insight to deformations away from the free fermionic point. 
Orbifold constructions would benefit as well, since the free fermionic
approach is suitable for a classification of models \cite{gkr,fknr}.
One can then envision a one--to--one map between the string vacuum
in the fermionic and orbifold representations. Such a map
will be instrumental in elucidating the phenomenological and cosmological
properties of specific string vacua in this class.

Moreover, we have
seen that, in general, non factorisable compactifications are promising
in their own right. For instance, by modifying the set of Wilson lines
in our construction, it should be not too difficult to find a three
generation standard like model for which matter is naturally embedded
into 16 dimensional representations of an underlying SO(10). That
guaranties a correct prediction for the hypercharges. Such a model
should then undergo further tests checking its relevance for the
description of the real world. 

Our investigations clearly show that ${\mathbb Z}_2 \times {\mathbb
Z}_2$ orbifolds of non factorisable six-tori should be well received
into the class of phenomenologically promising string vacua. 

\bigskip
\medskip
\leftline{\large\bf Acknowledgements}
\medskip

We would like to thank David Grellscheid, Thomas Mohaupt, Hans Peter 
Nilles, Sa\'{u}l
Ramos S\'{a}nchez, Patrick Vaudrevange and Ak{\i}n Wingerter for 
useful discussions. 
This work was supported by the PPARC and the University of Liverpool.


\vfill\eject

\bigskip
\medskip

\bibliographystyle{unsrt}

\vfill\eject
\end{document}